\documentclass[10pt,journal,compsoc]{IEEEtran}

\IEEEoverridecommandlockouts
\usepackage{graphicx}

\ifCLASSOPTIONcompsoc
    \usepackage[caption=false, font=normalsize, labelfont=sf, textfont=sf]{subfig}
\else
\usepackage[caption=false, font=footnotesize]{subfig}
\fi

\usepackage{graphicx}
\usepackage{cite}
\usepackage{amsfonts}
\usepackage{amsmath}
\usepackage{amsxtra}
\usepackage{amsbsy}
\usepackage{floatflt}
\usepackage{amssymb}
\usepackage{tabularx}
\usepackage{multirow}
\usepackage{nth}
\usepackage{comment}
\usepackage{epstopdf}
\usepackage{hhline}
\usepackage{siunitx}
\usepackage{booktabs}
\usepackage[bottom]{footmisc}
\usepackage[acronym]{glossaries}
\glsdisablehyper
\usepackage{upgreek}
\usepackage{algorithm}
\usepackage{algorithmicx,algpseudocode}
\usepackage{algpseudocode}
\usepackage{bm}
\usepackage[utf8]{inputenc}
\usepackage{lipsum}
\usepackage{array} 
\usepackage{epsfig}
\usepackage{times}
\usepackage{framed}
\usepackage{enumitem}
\usepackage{float}
\usepackage{etoolbox}
\usepackage{lipsum}
\usepackage{soul}
\usepackage{mathtools}
\usepackage{bbm}
\usepackage{makecell}
\usepackage{pifont}
\usepackage[export]{adjustbox} 
\usepackage{latexsym}
\usepackage{multicol}
\usepackage{eurosym}
\usepackage{xspace}
\usepackage{pdfpages}
\usepackage{verbatim}
\usepackage{stfloats}
\usepackage{blindtext}
\usepackage{colortbl}
\usepackage{xfrac}
\usepackage{relsize}

\newcommand{\ssub}[1]{{\scriptscriptstyle { #1}}}

\newcommand{\name}{COLoRIS}

\newcommand{\mb}[1]{\mathbf{{#1}}}
\newcommand{\tran}{\mathrm{T}}
\newcommand{\herm}{\mathrm{H}}
\newcommand{\diag}{\mathrm{diag}}
\newcommand{\tr}{\mathrm{tr}}
\DeclareMathOperator{\erf}{erf}

\newacronym{3d}{3D}{Three Dimensional}
\newacronym{ue}{UE}{User Equipment}
\newacronym[plural=RISs, firstplural=Reconfigurable Intelligent Surfaces (RISs)]{ris}{RIS}{Reconfigurable Intelligent Surface}
\newacronym{bs}{BS}{Base Station}
\newacronym{mimo}{MIMO}{Multiple-Input Multiple-Output}
\newacronym{miso}{MISO}{Multiple-Input Single-Output}
\newacronym{siso}{SISO}{Single-Input Single-Output}
\newacronym{upa}{UPA}{Uniform Planar Array}
\newacronym{ula}{ULA}{Uniform Linear Array}
\newacronym{aoa}{AoA}{Angle of Arrival}
\newacronym{aod}{AoD}{Angle of Departure}
\newacronym{snr}{SNR}{Signal to Noise Ratio}
\newacronym{kpi}{KPI}{Key Performance Indicator}
\newacronym{pdf}{PDF}{Probability Density Function}
\newacronym{pmf}{PMF}{Probability Mass Function}
\newacronym{fi}{FI}{Fisher Information}
\newacronym{crb}{CRB}{Cramér–Rao Bound}
\newacronym{csi}{CSI}{Channel State Information}
\newacronym{rv}{RV}{Random Variable}
\newacronym{pgm}{PGM}{Projected Gradient Method}
\newacronym{mse}{MSE}{Mean Squared Error}
\newacronym{ml}{ML}{Machine Learning}
\newacronym{mcu}{MCU}{Micro Controller Unit}
\newacronym{dnn}{DNN}{Deep Neural Network}
\newacronym{nlos}{NLoS}{non-Line of Sight}
\newacronym{los}{LoS}{Line of Sight}
\newacronym{soa}{SoA}{State-of-Art}
\newacronym{usart}{USART}{Universal Synchronous/Asynchronous Receiver/Transmitter}
\newacronym{usb}{USB}{Universal Serial Bus}
\newacronym{isac}{ISAC}{Integrated Sensing and Communication}
\newacronym{nn}{NN}{Neural Network}
\newacronym{slam}{SLAM}{Simultaneous Localization and Mapping}
\newacronym{iqr}{IQR}{Interquartile Range}
\newacronym{hpbw}{HPBW}{half-power beamwidth}
\newacronym{tdoa}{TDOA}{time difference of arrival}

\begin{document}

\title{\name{}: Localization-agnostic Smart Surfaces Enabling Opportunistic ISAC in 6G Networks}

\author{Guillermo Encinas-Lago,~\IEEEmembership{Student Member,~IEEE,}
Francesco Devoti,~\IEEEmembership{Member,~IEEE,} \\
Marco Rossanese,~\IEEEmembership{Student Member,~IEEE,}
Vincenzo~Sciancalepore,~\IEEEmembership{Senior Member,~IEEE,}\\
Marco Di Renzo,~\IEEEmembership{Fellow,~IEEE,}
Xavier~Costa-P\'erez,~\IEEEmembership{Senior Member,~IEEE,}

\thanks{\textit{Guillermo Encinas-Lago is with NEC Laboratories Europe, 69115 Heidelberg, Germany; Universit\'e Paris-Saclay, CNRS, CentraleSup\'elec, Laboratoire des Signaux et Syst\`emes, 3 Rue Joliot-Curie, 91192 Gif-sur-Yvette, France; and i2Cat, 08034 Barcelona, Spain. 
Francesco Devoti, Marco Rossanese and Vincenzo Sciancalepore are with NEC Laboratories Europe. 
Marco Di Renzo is with Universit\'e Paris-Saclay, CNRS, CentraleSup\'elec, Laboratoire des Signaux et Syst\`emes. (marco.di-renzo@universite-paris-saclay.fr), and with King's College London, Centre for Telecommunications Research -- Department of Engineering, WC2R 2LS London, United Kingdom (marco.di\_renzo@kcl.ac.uk). Xavier Costa-P\'erez is with i2Cat, ICREA, and NEC Laboratories Europe.
This work was partially supported by the Smart Networks and Services Joint Undertaking (SNS JU) Horizon Europe project under Grant Agreement 101139130 (6G-DISAC), by the SNS JU Horizon Europe Project under Grant Agreement No. 101192521 (MultiX), and by the SNS JU Horizon Europe Project under Grant Agreement No. 101139161 (INSTINCT). Views and opinions expressed are however those of the authors only and do not necessarily reflect those of the European Union or SNS JU. Neither the European Union nor the granting authority can be held responsible for them. The work of M. Di Renzo was also supported in part by the Agence Nationale de la Recherche (ANR) through the France 2030 project ANR-PEPR Networks of the Future under grant agreement NF-PERSEUS 22-PEFT-004, and by the CHIST-ERA project PASSIONATE under grant agreements CHIST-ERA-22-WAI-04 and ANR-23-CHR4-0003-01. \newline
Email of the corresponding author: guillermo.encinas@i2cat.net.}}%
}

This work has been submitted to the IEEE for possible publication. Copyright may be transferred without notice, after which this version may no longer be accessible. \newpage

\maketitle

\begin{abstract}
The integration of Smart Surfaces in 6G communication networks, also dubbed as \glsunset{ris}\acrlongpl{ris} (\glspl{ris}), is a promising paradigm change gaining significant attention given its disruptive features. \glspl{ris} are a key enabler in the realm of 6G \gls{isac} systems where novel services can be offered together with the future mobile networks communication capabilities.
This paper addresses the critical challenge of precisely localizing users within a communication network by leveraging the controlled-reflective properties of \gls{ris} elements without relying on more power-hungry traditional methods, e.g., GPS, adverting the need of deploying additional infrastructure and even avoiding interfering with communication efforts. Moreover, we go one step beyond: we build \name{}, an \emph{Opportunistic ISAC} approach that leverages localization-agnostic \gls{ris} configurations to accurately position mobile users via trained learning models. 
Extensive experimental validation and simulations in large-scale synthetic scenarios show $\textbf{5\%}$ positioning errors (with respect to field size) under different conditions.
Further, we show that a low-complexity version running in a limited off-the-shelf (embedded, low-power) system achieves positioning errors in the $\textbf{11\%}$ range at a negligible $\textbf{+2.7\%}$ energy expense with respect to the classical \gls{ris}. 
\end{abstract}


\glsresetall

\section{Introduction}
\label{sec:introduction}

In the ever-evolving landscape of modern technology, the implementation of emerging technology, such as \glspl{ris},\footnote{Note that the terms \gls{ris} and Smart Surface can be used interchangeably within the paper, as they refer to the same physical device.} has emerged as a transformative force, reshaping the way we perceive and interact with the propagation environment~\cite{DiRenzo2020_Jsac}. \gls{ris} allows directly controlling how electromagnetic waves propagate throughout the environment, opening up to never-explored use-cases. 

The merger of the sensing sphere with the communications world engenders the novel concept of \gls{isac}~\cite{ZRWGY_surveys_2022}. \gls{isac} represents a cutting-edge paradigm that heralds an era of interconnectedness and data-driven decision-making: At the heart of this evolution are smart surfaces, which act as key enablers, unleashing the full potential of \gls{isac} applications.
This holistic approach empowers systems not only to transmit and receive data, but also to collect, process, and utilize it for a wide array of applications, such as localization, detection, etc.~\cite{QLYA_Commag_2023}. \gls{isac}'s domain spans from smart cities and industrial automation to healthcare and environmental monitoring, offering solutions that are not only more efficient but also more environmentally sustainable. 
Nevertheless, the integration of such components presents a multitude of intricate challenges that pose a substantial threat to the overall system stability~\cite{gonzalez2024integrated}. Indeed, the network infrastructure that is currently deployed may necessitate a sophisticated orchestration and management framework to extend its functionality for data acquisition~\cite{YHLPZ_TSP_2022}, while still serving its conventional communication purposes. 

This calls for a compelling transition of the preexisting network equipment towards a novel and frictionless utilization, which inherently supports the extraction of contextual information for sophisticated machine learning models capable of capturing and deducing sensory data, going beyond the \gls{ris}-based passive radar applications~\cite{FKCWS_JSAC_2022} or \gls{slam} techniques~\cite{YWJ_JSAC_2022}. This innovative paradigm can be termed as \emph{opportunistic \gls{isac}} as it seamlessly capitalizes on the wireless network configurations to infer pertinent positioning information accurately. Unlike most localization systems leveraging \gls{ris}, our method does not interfere with the overall communication operations.
In particular, we do not alter how the \gls{ris} is used, configured, or optimized.
Indeed, we rely only on the available (selected) \gls{ris} configuration information for communication to, eventually, perform localization seamlessly.
Conversely, existing localization techniques with \gls{ris} require specific operations, e.g., time blocks exclusively devoted to localization~\cite{YHLPZ_TSP_2022}, dedicated configurations~\cite{FKCWS_JSAC_2022}, configuration \nocite{bjornson2019intelligent}sweeps~\cite{abu2021near}, etc., done \emph{ex-professo} to locate the user, that harm communication performance and compatibility with existing equipment.
This not only sets a new benchmark in efficiency by requiring significantly less power and computational resources, thereby heralding a sustainable and cost-effective approach. However, it also underscores the limitations of traditional analytical models, which often remain intractable when compared to the flexibility and scalability of machine learning-based methodologies.

Within this context, we pioneer a novel \gls{ml}-based framework, namely \emph{Configuration-based Opportunistic Localization via \gls{ris} (\name{})}, which harnesses successive \gls{ris} configurations, primarily focused on augmenting the overall system communication capabilities, to forecast exact user positions.

{\bf Contributions.} To summarize, \emph{(C1)} we introduce \name{}, a novel opportunistic \gls{isac} framework that exploits \gls{ris} configurations with an integrated error prediction mechanism, enabling precise user localization; \emph{(C2)} we demonstrate the practical feasibility of our framework through a comprehensive approach, including a \gls{fi}-based performance metric, machine learning implementation, and validation of sub-meter positioning accuracy in realistic scenarios; \emph{(C3)} we present a low-cost, energy-efficient prototype developed using off-the-shelf devices, implementing a complexity reduction strategy and validating the system's capability to operate using minimal power, with the potential to perform tens of thousands of daily position measurements on a small coin battery.
These contributions collectively advance the state-of-the-art in \gls{ris}-based localization technologies for emerging 6G networks.

{\bf Notation.} Matrices are in bold capital letters ($\mb{X}$), vectors are in small capital letters ($\mb{x}$), and scalars are in small letters ($x$). $\angle (\cdot)$ denotes the unary angle operator. The $n$-th element of the vector $\mb{x}$ is denoted as $\{\mb{x}\}_{n}$. Similarly, the $m$-th column of the matrix $\mb{X}$ is denoted as $\{\mb{X}\}_{m}$. The transpose, the Hermitian, and the trace of the matrix $\mb{X}$ are denoted as $\mb{X}^\tran$, $\mb{X}^\herm$, and $\tr(\mb{X})$, respectively. The norm of a vector $\mb{x}$ is $\|\mb{x}\|$. The gradient operator with respect to the $N\times1$ vector $\mb{x}$ is $\nabla_{\mb{x}}= [\frac{\partial}{\partial x_1}, \dots, \frac{\partial}{\partial x_{\ssub{N}}}]^\tran$. The expected value of a function $f(Z)$ is $\operatorname{\mathbb{E}}\left[f(Z)\right]$, and the average of a quantity $x$ is denoted as $ \overline{x}$. Sets are in calligraphic uppercase letters $(\mathcal{X})$, and their cardinality is denoted as $|\mathcal{X}|$.

\section{Analysis}
\label{sec:system_model}
\begin{figure}[t]
        \center
        \includegraphics[width=.3569\textwidth, trim = {0cm .5cm 0cm 0cm}]{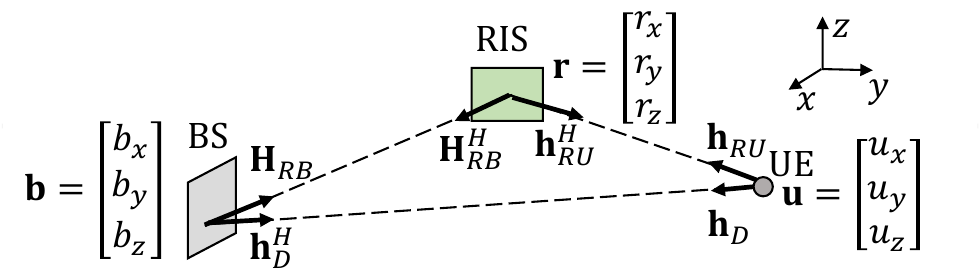}
        \caption{\label{fig:system_model}
        Geometrical representation of the considered scenario, including the \gls{bs} position $\mb{b}$, the \gls{ris} position $\mb{r}$, and a \gls{ue} position $\mb{u}$.}
\end{figure}

We consider a reference scenario as depicted in Fig.~\ref{fig:system_model} with a \gls{ris} deployed within the service area of a \gls{bs} serving a single antenna \gls{ue}. The \gls{bs} is equipped with an $M$ elements antenna array.
The \gls{ris} comprises $N = N_y N_z$ elements distributed as an \gls{upa} lying on the $yz$-plane, where $N_y$ and $N_z$ are the number of elements on the $y$- and the $z$-axis, respectively.
The \gls{bs} and the \gls{ris} antenna arrays are centered in $\mb{b}$, and $\mb{r}$, respectively, while the antenna of the \gls{ue} is centered in $\mb{u}$.
The corresponding \gls{miso} channel is defined as follows
\begin{align}
\label{eq:channel}
\mb{h}^{\herm} \triangleq  \mb{h}_{\ssub{RU}}^\herm \mb{\Theta}^\herm \mb{H}_{\ssub{RB}} + \mb{h}_{\ssub{D}}^\herm  \in \mathbb{C}^{M \times 1},
\end{align}
with $\mb{h}_{\ssub{D}} \in \mathbb{C}^{M \times 1}$ denoting the direct \gls{bs}-\gls{ue} channel, $\mb{H}_{\ssub{RB}} \in \mathbb{C}^{N \times M}$ and $\mb{h}_{\ssub{RU}} \in \mathbb{C}^{N \times 1}$ corresponding to the \gls{ris}-\gls{bs} and the \gls{ris}-\glspl{ue} paths, respectively, and $\mb{\Theta} \in \mathbb{C}^{N \times N}$ is the \gls{ris} configuration that is described in its  most generic form as
\begin{align}
    \diag{(\mb{\Theta}}) = [\alpha_1 e^{j\theta_1}, \dots, \alpha_{\ssub{N}} e^{j\theta_{\ssub{N}}} ]^\tran,
\end{align}
where $\alpha_n \in [0,1]$ and $\theta_n \in [0,2\pi)$ are the gain and the phase shift of the $n$-th element of the \gls{ris}, respectively. We consider the \gls{ris} device to be capable of phase modifications i.e.,  $\alpha_n = 1 \,\, \forall \,\, n = 1, \dots, N$, and $ \diag{(\mb{\Theta}}) = [e^{j\theta_1}, \dots, e^{j\theta_{\ssub{N}}} ]^\tran$.

We denote the \gls{ris} array response with the vector $\mb{a}_{\ssub{R}}(\mb{p}) \in \mathbb{C}^{N \times 1}$ with elements defined as the following
\begin{align}
    \{\mb{a}_{\ssub{R}}(\mb{p})\}_n \triangleq G(\mb{p}) e^{ -\jmath \frac{2 \pi }{\lambda} \left(\Vert \mb{p}-\mb{q}_n\Vert-\Vert \mb{p}-\mb{p}_{\text{ref}}\Vert \right)}, \label{eq:ris_response}
\end{align}
where $\mb{p} \in \mathbb{R}^3$ is the point of departure or arrival of the signal, $\lambda$ is the wavelength at the carrier frequency $f_0$, $\mb{q}_n=\mb{p}_n-\mb{p}_{\text{ref}}  \in \mathbb{R}^3$ is the offset between the absolute position $\mb{p}_n$ of the $n$-th \gls{ris} element and the arbitrary \gls{ris} reference point $\mb{p}_{\text{ref}} \in \mathbb{R}^3$, and $G(\mb{p})$ is the gain of the element in the direction of $\mb{p}$, which is $G(\mb{p}) = 1$ under the assumption of patch antenna elements.
Importantly, Eq.~\eqref{eq:ris_response} accounts for the wavefront curvature at the \gls{ris} as it depends directly on the location $\mb{p}$, rather than on angles of arrival and departure only~\cite{abu2021near, rahal2021ris}. The \gls{bs} array response vector $\mb{a}_{\ssub{B}}(\bm{p}) \in \mathbb{C}^{M \times 1}$ is defined similarly.

Having defined the array steering vectors, we can now define the channels $\mb{H}_{\ssub{RB}}$, $\mb{h}_{\ssub{RU}}$, and $\mb{h}_{\ssub{D}}$ as follows

\begin{align}
    \{\{\mb{H}_{\ssub{RB}}\}_{m}\}_n &\triangleq \sqrt{\gamma(\mb{p}_n,\mb{p}_m)}\{\mb{a}_{\ssub{R}}(\mb{p}_m)\}_n\{\mb{a}^{\herm}_{\ssub{B}}(\mb{p}_n)\}_m, \label{eq:channel_bs_ris}\\
    \{\mb{h}_{\ssub{RU}}\}_n &= \{\mb{h}_{\ssub{RU}}(\mb{u})\}_n  \triangleq \sqrt{\gamma(\mb{p}_n,\mb{u})}\{\mb{a}_{\ssub{R}}(\mb{u})\}_n, \label{eq:channel_ris_ue}\\
    \{\mb{h}_{\ssub{D}}\}_m &= \{\mb{h}_{\ssub{D}}(\mb{u})\}_m  \triangleq \sqrt{\gamma(\mb{p}_m,\mb{u})}\{\mb{a}_{\ssub{B}}(\mb{u})\}_m, \label{eq:channel_bs_ue}
\end{align}
\noindent $m=1\dots M$, $\gamma(\mb{x},\mb{y})$ is the path gain between two given locations $\mb{x}$, $\mb{y} \in \mathbb{R}^3$ and is defined as the following
\begin{align}
\gamma(\mb{x},\mb{y}) \triangleq \gamma_0 \left( \tfrac{d_0}{\|\mb{x} - \mb{y}\|} \right)^\beta,
\label{eq:path_gain}
\end{align}
where $\gamma_0$ is the channel power gain at a reference distance $d_0$ and $\beta$ is the pathloss exponent.

We consider a generic, non-ideal, channel estimation process that returns the estimation of $\hat{\mb{H}}_{\ssub{RB}}$, $\hat{\mb{h}}_{\ssub{RU}}(\mb{u})$, and $\hat{\mb{h}}_{\ssub{D}}(\mb{u})$ as \gls{csi}.
Moreover, we assume the presence of a phase selection agent that optimizes the \gls{ris} configuration based on the available \gls{csi}.
To ease the notation, we introduce an alternative but equivalent representation of $\mb{\Theta}$ with the definition of the following vector
\begin{align}
\boldsymbol{\uptheta} = \diag{ (\angle \mb{\Theta})} = \left[\theta_1, \theta_2, \dots, \theta_{\ssub{N}} \right]^\tran \in \{0,2\pi\}^{N \times 1}.
\end{align}
We denote the selected \gls{ris} configuration as $\bar{\boldsymbol{\uptheta}}$, and we describe the optimization process executed by the agent as 
\begin{align}
\bar{\boldsymbol{\uptheta}} =\angle \bar{\mb{\Theta}}&= g(\hat{\mb{H}}_{\ssub{RB}}, \hat{\mb{h}}_{\ssub{RU}}, \hat{\mb{h}}_{\ssub{D}}) = \underbrace{g(\mb{H}_{\ssub{RB}}, \mb{h}_{\ssub{RU}}, \mb{h}_{\ssub{D}})}_{\mb{\uptheta}^*(\mb{u})} + \mb{e}_{\uptheta}, \label{eq:ris_configuration}
\end{align}
where $g(\cdot)$ is the \gls{ris} configuration optimization process, $\mb{\uptheta}^*(\mb{u})$ is the optimal configuration obtained when perfect \gls{csi} is available, which depends on the \gls{ue} position $\mb{u}$ due to \eqref{eq:channel_ris_ue} and \eqref{eq:channel_bs_ue}, and $\mb{e}_{\uptheta}$ is the error introduced by the non-ideal channel estimation process.
Note that phase shift values can be either discrete or continuous, depending on the \gls{ris} design~\cite{fara2022prototype, rossanese2022designing}. We characterize $\bar{\boldsymbol{\uptheta}}$ in the continuous case, and face the discrete case later in this section.

In its most general form---as detailed in Section~\ref{sec:configuration_methods}---the configuring agent operates as a phase measurement device that extracts the phase of the \gls{csi}.
Given the direct dependence of \gls{csi} quality on the \gls{snr},\footnote{\gls{csi} estimation techniques (as pilot signals) are impacted by \gls{snr}: higher \gls{snr} improves accuracy and vice versa \cite{steven1993fundamentals}.}
we assume the spread of the error $\mb{e}_{\uptheta}$ to be inversely proportional to the \gls{snr}.
Furthermore, for the sake of tractability and without loss of generality, we assume phase measurements to be independent at each \gls{ris} element.
Hence, we model the error $\mb{e}_{\Theta}$ as a normal \gls{rv} with distribution  $\mb{e}_{\uptheta}\sim\mathcal{N}(\mb{0}_{\ssub{N}}, \sigma_{\theta}^2\mb{I}_{\ssub{N}})$, with $\sigma^2_\theta = \frac{1}{2\gamma}$, where $\gamma$ denotes the \gls{snr}.
Accordingly, $\bar{\boldsymbol{\uptheta}}$ is a normal \gls{rv} with distribution

\begin{align}
\bar{\boldsymbol{\uptheta}}\sim f_{\bar{\boldsymbol{\uptheta}}}(\bar{\boldsymbol{\uptheta}};\mb{u}) = \mathcal{N}(\boldsymbol{\uptheta}^*(\mb{u}), \sigma^2_{\uptheta}\mb{I}_{\ssub{N}}). \label{eq:theta_distribution}
\end{align}
Section~\ref{sec:proof_of_gaussian_assumption} provides a formal demonstration of Eq.~\eqref{eq:theta_distribution}.

When considering an \gls{ris} hardware capable of discrete phase shifts, we assume the set of feasible configurations 
\begin{equation}
    \mathcal{Q} = \left\{\Delta m : m = 0,\dots, 2^{Q-1}, m \in \mathbb{N} \right\}, \label{eq:phase_quantization_set}
\end{equation}
 where $Q$ denotes the phase shifts quantization level, and $\Delta = \tfrac{2\pi}{2^Q}$. The total number of feasible configurations is $| \mathcal{Q} |=2^{Q}$. Phase shifts are set according to the quantization function $q: \mathbb{R} \rightarrow \mathcal{Q}$ defined as
\begin{equation}
q(x) = \Delta \left\lfloor \tfrac{x}{\Delta} + \tfrac{1}{2} \right\rfloor.
\end{equation}
The quantized \gls{ris} configuration is obtained as $\bar{\boldsymbol{\uptheta}}_{\ssub{Q}} = q(\bar{\boldsymbol{\uptheta}})$. Again, $\bar{\boldsymbol{\uptheta}}_{\ssub{Q}}$ is a \gls{rv}, which is distributed as the following
\begin{align}
\bar{\boldsymbol{\uptheta}}_{\ssub{Q}}\sim f_{\bar{\boldsymbol{\uptheta}}_{\ssub{Q}}}(\bar{\boldsymbol{\uptheta}}_{\ssub{Q}};\mb{u}) =  \prod_{n=1}^{N} f_{\bar{\theta}_{\ssub{Q},n;\mb{u}}}(\bar{\theta}_{\ssub{Q},n};\mb{u}), \label{eq:theta_distribution_quant}
\end{align}
with
\begin{align}
f_{\bar{\theta}_{\ssub{Q},n}}\!(\bar{\theta}_{\ssub{Q},n};\mb{u}) \! = \!\! \sum_{m=0}^{2^{Q-1}} \!P\!\left(\bar{\theta}_{\ssub{Q},n}\!=\! \Delta m;\mb{u}\right) 
\!\delta\! \left(\bar{\theta}_{\ssub{Q},n} \!-\! \tfrac{2\pi}{2^Q} m \right)\!,\label{eq:theta_distribution_quant_el}
\end{align}
where $\delta(\cdot)$ is the delta function, and
\begin{equation}
P\left(\bar{\theta}_{\ssub{Q},n}= \Delta m;\mb{u}\right) =  \int_{\Delta(m-0.5)}^{\Delta(m+0.5)} f_{\bar{\theta}_{n}}(\theta;\mb{u}) d\theta\label{eq:P}
\end{equation}
is the probability of selecting $\bar{\theta}_{\ssub{Q},n}= \Delta m$.

\subsection{Optimizing RIS Configurations}
\label{sec:configuration_methods}
The \gls{ris} configuration process is generally complex, as it implies the joint optimization of the \gls{bs} precoder and the \gls{ris} configuration to maximize a given objective \glspl{kpi}, and is currently still an open problem~\cite{pan2022overview}. Hereafter, we summarize the main joint optimization strategies: $i$) \emph{Gradient optimization}: iterative adjustment of \gls{ris} configuration and \gls{bs} to optimize given metrics (e.g., \gls{snr}) in the direction of the gradient; $ii$) \emph{Alternating optimization}: optimization of each independent subproblem while keeping the other fixed~\cite{mursia2020risma,yu2020robust}; $iii$) \emph{\acrlong{ml}}: application of \gls{ml} agents to face \gls{ris} configuration problems~\cite{faisal2022machine,encinas2023unlocking}; $iv$) \emph{Coherent paths}: a practical approach that maximizes the gain of the reflected path while achieving constructive interference with the direct path~\cite{bjornson2019intelligent}.

\begin{figure}[t]
\centering
\includegraphics[width=1\linewidth, trim = {0cm 0.75cm 0cm 0cm}]{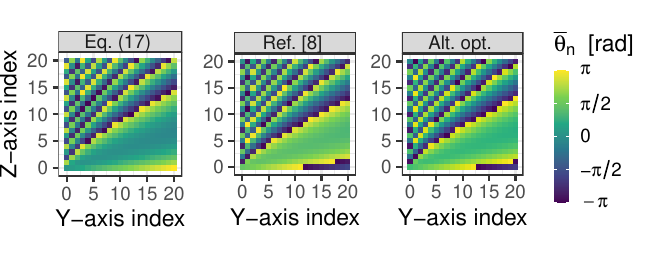}
\caption{\gls{ris} configurations from coherent paths optimization Eq.~\eqref{eq:coherent_solution}, optimization method from~\cite{bjornson2019intelligent} and alternate optimization algorithm from~\cite{mursia2020risma}.
} \label{fig:coherent_vs_risma}
\end{figure}

Considering the coherent paths configuration solution, the \gls{ris} configuration process $g(\cdot)$ can be written in a closed form while still obtaining results comparable to more complex optimization procedures~\cite{albanese2022marisa}.
We anticipate that this optimization approach enables us to create offline ground-truth data that we use for training \name{}, as described in Section~\ref{sec:localization_method}.

Let us consider the rate $R$ maximization problem in \gls{miso} scenario, which can be formulated as follows
\begin{align}
\max_{\bar{\mb{\Theta}}} R = \log_2 \left( 1 + \tfrac{P_t}{\sigma^2}\left| \mb{w}^{\herm} \left(\mb{H}_{\ssub{RB}}^\herm \bar{\mb{\Theta}} \mb{h}_{\ssub{RU}} + \mb{h}_{\ssub{D}} \right)\right|^2 \right), \label{eq:rate}
\end{align}
where $P_t$ is the transmission power, and $w \in \mathbb{C}^{M\times1} $ is the \gls{bs} precoder.  Following~\cite{albanese2022marisa}, Eq.~\eqref{eq:rate} is maximized when
\begin{equation}
\bar{\boldsymbol{\uptheta}} = \angle z_{\ssub{D}} - \angle \mb{z}_{\ssub{BRU}}, \label{eq:coherent_solution}
\end{equation}
where $\mb{z}_{\ssub{BRU}} = \mb{h}_{\ssub{RU}}^* \circ \mb{H}_{\ssub{BR}}\mb{w}$, with $\circ$ denoting the Hadamard product, and $z_{\ssub{D}}=\mb{w}^\herm\mb{h}_{\ssub{D}}$ depend only on the reflected and on the direct path components, respectively.
The so-obtained \gls{ris} configuration maximizes the reflected path gain while achieving phase alignment between the direct and reflected paths. A similar solution is found in~\cite{bjornson2019intelligent} for the \gls{siso} case. 

Finally, an example of \gls{ris} configuration optimized with Eq.~\eqref{eq:coherent_solution}, the solution proposed in~\cite{bjornson2019intelligent} and the more complex alternate optimization method proposed in~\cite{mursia2020risma} is shown in Fig.~\ref{fig:coherent_vs_risma}.
The strong similarity between the two output configurations validates our initial hypothesis: while the usage of a \gls{ml} architecture instead of an analytical approach makes our proposal less dependant on the detail of the configuration method employed, \emph{the similarity of the configurations obtained with different solution methods reinforces the validity of \name{} across them}.
Since finding, producing, and approximating an optimal \gls{ris} configuration is a well-studied topic, our work relies on the assumption that \glspl{ris} devices are already configured with existing solutions and we build upon those.

\subsection{Localization Primer}
\label{sec:pillars}

\begin{figure}[t]
\centering
\includegraphics[width=1\linewidth, trim = {0cm .75cm 0cm 0cm}]{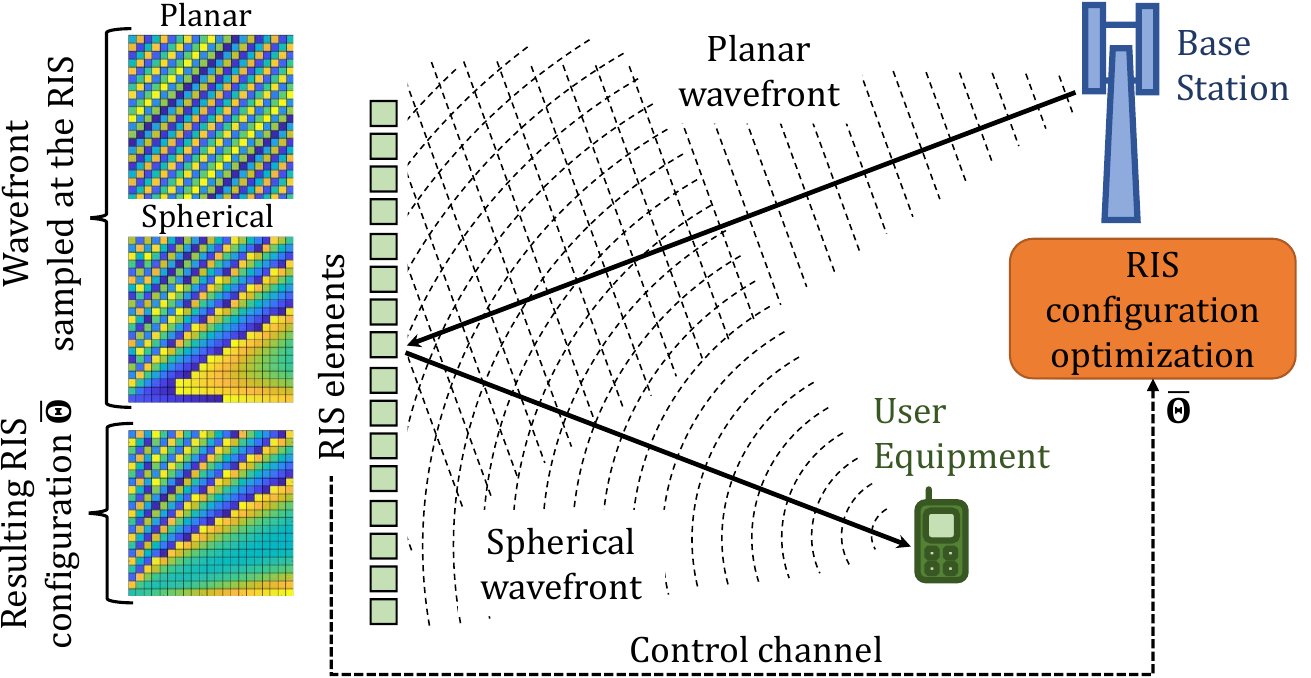}
\caption{Example of communication scenario with \gls{bs} in the far-field and \gls{ue} in near-field regimes of the \gls{ris}, with a representation of the corresponding wavefront spatially sampled at \gls{ris} elements, and resulting \gls{ris} configuration. \label{fig:scenario_with_wavefronts} }
\end{figure}

The \gls{ris} configuration is linked with the channel conditions between \gls{bs}, \gls{ris} and \gls{ue} (Eq.~\eqref{eq:ris_configuration}).
Such channels reflect the wavefront describing the signal propagation between devices, which depends on their relative position (Eq.~\eqref{eq:ris_response}).
To better clarify, we consider the reference scenario in Fig.~\ref{fig:scenario_with_wavefronts}, wherein the \gls{ue} and the \gls{bs} are in the near-field and the far-field regimes of the \gls{ris}, respectively.
Consequently, at the \gls{ris} site, the signal from the \gls{ue} forms a spherical wavefront whereas the signal from the \gls{bs} creates a planar one. The planar wavefront is normal to the signal direction between the \gls{bs} and the \gls{ris}, while the curvature of the spherical one depends on both the \gls{ue}-\gls{ris} signal direction and the distance among them.

The selected \gls{ris} configuration depends on the phase distribution pattern of the impinging wavefronts. This dependence is clear from Eq.~\eqref{eq:coherent_solution}, where $\bar{\boldsymbol{\uptheta}}$ is a function of the direct path ($z_{\ssub{D}}$) and the reflected path ($\mb{z}_{\ssub{BRU}}$). While $z_{\ssub{D}}$ aligns the phases of the direct and reflected paths, $\mb{z}_{\ssub{BRU}}$ matches the phase shift pattern on the surface $\bar{\boldsymbol{\uptheta}}$ with the wavefront of $\mb{h}_{\ssub{RU}}$ and $\mb{H}_{\ssub{BR}}$.

Notably, $\mb{h}_{\ssub{RU}}$ depends on the \gls{ue} position $\mb{u}$, while $\mb{H}_{\ssub{BR}}$ on the \gls{bs} position. Hence, in principle, estimate $\mb{u}$ by the observation of $\bar{\boldsymbol{\uptheta}}$ is feasible. An estimator $\hat{\mb{u}}(\bar{\boldsymbol{\uptheta}})$ of the \gls{ue} location can be defined based on minimizing the norm of the difference between the controller-selected configuration and the one obtained with an approximation based on information available at the \gls{ris} as
\begin{equation}
 \hat{\mb{u}} = \hat{\mb{u}}(\bar{\boldsymbol{\uptheta}}) = \underset{\mb{u}}{\operatorname{argmin}} \| \operatorname{angdiff}(\tilde{g}(\mb{u}),\bar{\boldsymbol{\uptheta}}) \|,
\end{equation}
where $\tilde{g}(\cdot)$ approximates the \gls{ris} configuration process, and $\operatorname{angdiff}(a,b) = |((a-b+\pi) \operatorname{mod} 2\pi ) -\pi|$ returns the difference between two angles.

A crucial discussion arises within this context. To define an estimator $\hat{\mb{u}}(\bar{\boldsymbol{\uptheta}})$, it is imperative to possess prior knowledge of either the channel conditions or the relative positions of the \gls{bs} and the \gls{ris}. Indeed, while \gls{csi} are inaccessible at the \gls{ris} due to limited hardware capabilities, they can be approximated using models when information on the device deployment is available.
Additionally, a reliable approximation of the \gls{ris} configuration process run at the \gls{ris} controller is required directly at the \gls{ris} itself, which is typically unavailable due to the third-party nature of the controller.

It is worth noting that most \gls{ris} configuration processes assume prior knowledge of the locations of the \gls{bs} and \gls{ue}.
Despite in practice, this information is frequently imprecise or absent, to evaluate the feasibility of this localization approach, we assume that some information on the device deployment is available and that the controller optimizes the \gls{ris} while leveraging non-perfect \gls{csi}.
It is worth mentioning that the election of far-field for the \gls{bs} for this exposition is purely for the sake of clarity. In general, the \gls{bs} can be either in near- or far-field, without this constituting an assumption, limitation, or imposition of our work. However, to completely locate the \gls{ue} with a single \gls{ris} and no additional information, the \gls{ue} must be located necessarily in the near-field of the \gls{ris}, as only relative direction is available at far-field. Our method is at any rate capable of obtaining this information, as shown in Section~\ref{sec:results}, where it successfully works in near-field, producing accurate localization, and in far-field, producing accurate \gls{aoa}. Nonetheless, near-field conditions at the \gls{ris} will be common due to the trend towards larger \glspl{ris} operating at higher frequencies~\cite{jian2022reconfigurable},
which significantly increases the Fraunhofer distance.

\subsection{Fisher Information and Cramér–Rao Bound}
\label{sec:fisher_information}
The \acrfull{fi} and the \gls{crb} are well-accepted means to statistically characterize an estimator~\cite{steven1993fundamentals}. In particular, the \gls{fi} provides a measure of the amount of information that an observable \acrfull{rv} $X$ carries about an unknown parameter $Y$, given the distribution $f_{(X)}(x;y)$. The \gls{crb} is a statistical bound of the achievable accuracy of an unbiased estimator predicting the value of the parameter $Y$ from the observation of $X$.

Here we want to estimate the position $\mb{u}$ of an \gls{ue} from the observed configuration of a \gls{ris} optimized to serve said device. Hence, the unknown $\mb{u}$ belongs to $\mathbb{R}^{D\times1}$ with $D=3$ in the 3D localization case or $D=2$ in the 2D localization case\,\footnote{Note that our analysis is carried out considering the 2D case for the sake of simplicity. However, it can be readily adapted to 3D.}.
On the contrary, the observations are the individual values of phase shifts that collectively form the selected \gls{ris} configuration, i.e., $\bar{\boldsymbol{\uptheta}}$. As detailed in Section~\ref{sec:system_model}, the underlying process selecting the \gls{ris} configuration is subjected to noise, and ultimately, error.

We aim to estimate the unknown $\mb{u}$ from the observation of $\bar{\boldsymbol{\uptheta}}$, which is described by the \gls{pdf} $f_{\bar{\boldsymbol{\uptheta}}}(\bar{\boldsymbol{\uptheta}}; \mb{u})$. The \gls{fi} is denoted by the symmetric matrix $\mb{J}(\mb{u}) \in \mathbb{R}^{D \times D}$ defined as
\begin{equation}  \label{eq:fisher_info}
\mb{J}(\mb{u}) = 
\mathbb{E}\left[\left(
\nabla_{\mb{u}} \ln f_{\bar{\boldsymbol{\uptheta}}}(\bar{\boldsymbol{\uptheta}}; \mb{u})\right) \left(
\nabla_{\mb{u}} \ln  f_{\bar{\boldsymbol{\uptheta}}}(\bar{\boldsymbol{\uptheta}}; \mb{u})\right)^{\herm} \right],
\end{equation}
where $\ln(\cdot)$ is the natural logarithm, \cite{Zhao_Guibas_2003}.
In this context, high values of \gls{fi} entries indicate that the \gls{pdf} $f_{\bar{\boldsymbol{\uptheta}}}(\bar{\boldsymbol{\uptheta}}; \mb{u})$ is changing substantially depending on the user location $\mb{u}$. In other words, the observation of $\bar{\boldsymbol{\uptheta}}$ carries significant information on $\mb{u}$.
Vice versa, lower values of \gls{fi} entries indicate that the observation of $\bar{\boldsymbol{\uptheta}}$ carries little information on $\mb{u}$.
Consequently, the \gls{fi} allows us to evaluate the feasibility of the localization task via the observation of \gls{ris} configuration.

As described in Section~\ref{sec:system_model}, the selected configuration is independent at each \gls{ris} element. Hence, the \gls{pdf} of $\bar{\boldsymbol{\uptheta}}$ can be rewritten as $f_{\bar{\boldsymbol{\uptheta}}}(\bar{\boldsymbol{\uptheta}}; \mb{u}) = \prod_{n=1}^N f_{\bar{\theta}_n}(\bar{\theta}_n; \mb{u})$.
This, thanks to the product logarithmic identity,
allows us to rewrite Eq.~\eqref{eq:fisher_info} as $\mb{J}(\mb{u}) = \sum_{n=1}^N \mb{J}_{n}(\mb{u})$, with
\begin{equation} 
\mb{J}_{n}(\mb{u}) =
\mathbb{E}\left[\left(
\nabla_{\mb{u}} \ln  f_{\bar{\theta_n}}(\bar{\theta}_n; \mb{u})\right) \left(
\nabla_{\mb{u}} \ln  f_{\bar{\theta_n}}(\bar{\theta}_n; \mb{u})\right)^{\herm} \right]. \label{eq:fisher_info_2}
\end{equation}
Note that $\mb{J}_{n}(\mb{u})$ is the \gls{fi} matrix related to the single $n$-th \gls{ris} element.
We can now evaluate the amount of information on $\mb{u}$ that is carried by each configured \gls{ris} element individually,\footnote{This formulation allows us to treat each \gls{ris} element independently, rather than as part of a single-sensor model defined by beam width, e.g., ~\cite{liang2024joint}. This granular perspective enables identifying the most informative elements for our machine learning model while significantly reducing computational complexity as we detail in Section~\ref{sec:complexity_reduction_approach}.} which, as per~\cite{alshunnar2010comparison}, is the following
\begin{equation} \label{eq:info_per_element}
 I_n = \tr\left(\mb{J}_{n}(\mb{u})\right).
\end{equation}
We can use $I_n$ to evaluate the average \gls{fi} per element, i.e., $I = \frac{1}{N}\sum_{n=1}^N I_n$, which is a particularly useful \gls{kpi} to compare the theoretical performance of different \glspl{ris} layouts. It is worth noting that, thanks to the property of the trace operator, $I_n$ is invariable with rotations, hence independent of the reference system orientation.

It is worth pointing out that Eqs.~\eqref{eq:fisher_info} and~\eqref{eq:fisher_info_2} also apply in the case of discretized phase selection. Interestingly, in the case of $Q=1$, Eq.~\eqref{eq:fisher_info_2} boils down to the following
\begin{equation}
\mb{J}_{n}(\mb{u}) \! = \! \Delta^2 \!
\left(
\nabla_{\mb{u}} \ln  \! P\! \left(\bar{\theta}_{\ssub{Q},n} \! = \! \Delta \right) \!\right)\! \left(
\nabla_{\mb{u}} \! \ln \! P\left(\bar{\theta}_{\ssub{Q},n} \! = \!\Delta \right) \!\right)^{\herm}\!\!, \label{eq:fisher_info_binary}
\end{equation}
which greatly simplifies the \gls{fi} matrix computation.

The \gls{crb} is the inverse of the \gls{fi} matrix, i.e., $\mb{J}^{-1}(\mb{u})$, and statistically lower bounds the error covariance matrix for any unbiased estimator of $\hat{\mb{u}}(\mb{\bar{\boldsymbol{\uptheta}}})$.
More specifically, we have that
\begin{equation} \label{eq:cramer_rao_bound}
\mathbb{E} \left[ \left(\hat{\mb{u}}(\mb{\bar{\boldsymbol{\uptheta}}})-\mb{u}\right) \left(\hat{\mb{u}}(\mb{\bar{\boldsymbol{\uptheta}}})-\mb{u} \right)^{\herm}  \right] \succeq \mb{J}^{-1}(\mb{u}).
\end{equation}
In other words, the $d$-th diagonal element of $\mb{J}^{-1}(\mb{u})$ lower bounds the \gls{mse} of any unbiased estimator of $\{\mb{u}\}_d$ from $\bar{\boldsymbol{\uptheta}}$, \cite{steven1993fundamentals}.
Hence, the achievable localization accuracy is
\begin{equation} \label{eq:localization_accuracy}
    \sigma_{\mb{u},\ssub{CRB}}(\mb{u}) = \sqrt{\tr{\left(\mb{J}^{-1}(\mb{u})\right)}}.
\end{equation}

Finally, we can use Eq.~\eqref{eq:localization_accuracy} to evaluate localization performance over an area $A$ (instead of a point) and measure the portion of $A$ achieving sufficient localization performance, as described in~\cite{encinaslago2025rilocoisacorientedaisolution}. To do so, we define a threshold $\sigma_{\mb{u},\text{min}}$, which is the minimum desired localization accuracy, and a comparison function that tends to $0$ when the accuracy is below $\sigma_{\mb{u},\text{min}}$, and to $1$ when the accuracy is above $\sigma_{\mb{u},\text{min}}$. To this aim, we select the sigmoid function $S(x,x_\text{thr}) = \frac{1}{1+e^{(x_\text{thr}-x)\beta}}$, where $\beta > 0$ is a parameter defining the sharpness of the sigmoid.
Thus, the portion of $A$ achieving the minimum localization performance is obtained as
\begin{equation} \label{eq:area_localization_accuracy_metric}
   M(A) = \frac{1}{C_{\ssub{N}}}\int_{\ssub{A}} S\left(\sigma_{\mb{u},\ssub{CRB}}(\mb{u}), \sigma_{\mb{u},\text{min}}\right) dA ,
\end{equation}
where $C_{\ssub{N}} = \int_{\ssub{A}} dA$ is the size of $A$.

\section{Opportunistic ISAC Framework}
\label{sec:localization_method}
\begin{figure}[t]
        \center
        \includegraphics[width=.47\textwidth, trim = {0cm .75cm 0cm 0cm}]{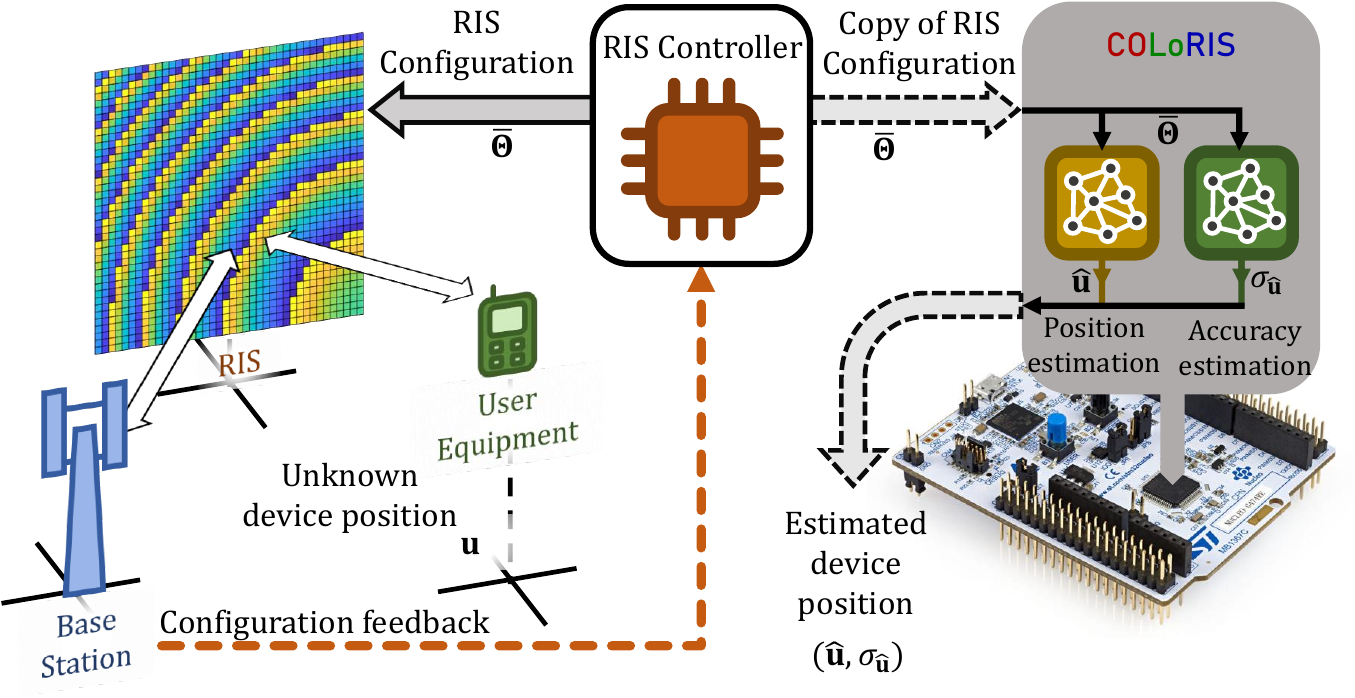}
        \caption{\label{fig:coloris_operations}
        Schematic representation of \name{}: the \gls{ris} configuration as input, the internal estimators, and the user position and accuracy as output.}
\end{figure}

We introduce \name{} that opportunistically leverages \gls{ris} configuration patterns to estimate the positions of served \glspl{ue}.
As discussed in Section~\ref{sec:pillars}, localizing \glspl{ue} based on deliberately triggered \gls{ris} configurations is inherently challenging due to limited \gls{ris} hardware capabilities, lack of \gls{csi} at the \gls{ris}, and the need for precise information about controller optimization operations. While an analytical approach to determine \gls{ue} positions is possible, the lack of knowledge about the specific optimization method used to select \gls{ris} configurations would limit its generality.
Hence, we rely on \gls{ml} methods and propose \name{}, \emph{an opportunistic framework that retrieves user positions with very high accuracy}.

\gls{ml} tools have been previously applied to address open challenges with \gls{ris} related problems, yielding promising results~\cite{encinas2023unlocking}.
Thanks to \gls{ml}, our method can automatically adapt to various propagation conditions and deployment scenarios without significant changes or external adaptations. Additionally, our design approach includes post-processing methods to greatly simplify its computational complexity (for more details please refer to Section~\ref{sec:results}). This makes it compatible with the limited hardware capabilities typically found in \gls{ris} prototypes~\cite{rossanese2022designing}, aligning with the cost-efficiency and low power consumption expected from such devices~\cite{prototype_TC21}.

The workflow of \name{} is depicted in Fig.~\ref{fig:coloris_operations}. In particular, we design our solution as a module running alongside the \gls{ris} aided communication system\footnote{
The design phase of \name{} does not strictly rely on how and where the \gls{ris} optimization is carried out, as it can occur at different entities~\cite{ComMag2021_strinati}.
}. The selected \gls{ris} configuration invariably passes through the \gls{ris} controller, which wields a privileged role in configuring the surface elements~\cite{ris2023etsi}. While \gls{bs}, \gls{ris} controller, and \gls{ris} devices establish and maintain their communications, the controller forwards in parallel the optimized \gls{ris} configuration to \name{}, which subsequently processes such information. Importantly, the localization process operates \textit{opportunistically} and does not disrupt the ongoing communication activities. At the same time, the simultaneous generation of position estimation and an associated accuracy facilitates the use of \name{} in cooperative localization and sensor fusion techniques~\cite{wymeersch2009cooperative, sasiadek2002sensor}.

This opportunistic design also enables \name{} to adapt seamlessly to dynamic environments where device mobility causes rapid changes in channel conditions. Such changes are managed by the specific \gls{ris} optimization method in use, which dynamically adapts the \gls{ris} configuration to the real-time channel conditions. Once a new \gls{ris} configuration becomes available, \name{} treats it as a new input, ensuring timely and accurate localization without requiring modifications to the core logic of the network or compromising its responsiveness.
Rapid inference capabilities of \name{} are crucial for responding quickly to new configurations. However, scalability in high-mobility scenarios depends on the relative durations of the \gls{ris} optimization, reconfiguration process, and location inference. When inference time is negligible, the localization process is primarily constrained by \gls{ris} reconfiguration. Conversely, if inference time exceeds reconfiguration time, only a subset of configurations can be processed during rapid mobility, potentially impacting localization performance.

\subsection{Machine Learning and Training Options}
\label{sec:ml_and_training_options}
The potential practical applications of such a device, which would operate within very heterogeneous situations and dynamic environments and the availability of large amounts of data through simulations, further justifies the choice of \gls{ml} techniques. In our scenario, we propose a straightforward approach: we use a \gls{dnn} to directly reconstruct the position of a \gls{ue} using the \gls{ris} configuration as input when optimized to serve said \gls{ue}.
The \gls{dnn} consists of an input layer with $N$ elements, which exploits the \gls{ris} configuration $\bar{\boldsymbol{\uptheta}}$.
One or more hidden layers---each with $N_h$ elements---allow the network to capture complex relationships and patterns in the input data.
An output layer with $D$ neurons, which provide as output the predicted location \gls{ue} $\hat{\mb{u}}$ in terms of coordinates (e.g., $\hat{u}_x$ and $\hat{u}_y$, when $D=2$). We select the hyperbolic tangent for the activation function in hidden layers, and the identity function for the output layer, corresponding to a linear regressor~\cite{gianola2011predicting}.
The second aspect of \name{} involves evaluating the precision of the localization process, ensuring its applicability in scenarios where accuracy determination is crucial, such as in systems that integrate position data from multiple sources, or in applications where meeting a specific accuracy benchmark is essential for the usability of the data.
Thus, we reuse two elements of the \gls{fi} and \gls{crb} analysis as shown in Section~\ref{sec:system_model}: $i$) the assumption of the availability of an unbiased estimator, and $ii$) the dependency of the achievable accuracy with the spatial gradient of the \gls{ris} configurations.

Let us consider the function $\bar{\boldsymbol{\uptheta}} = \mb{c}(\mb{u})$
that gives the \gls{ris} configuration serving an \gls{ue} at a given location.
If $\mb{c}(\mb{u})$ is known at \name{}, we can approximate the location estimation accuracy by computing its gradient around the estimated \gls{ue} location $\hat{\mb{u}}$. Hence, we can obtain an approximation of the accuracy by considering that
\begin{equation}
\sigma^2_{\mb{u}}(\hat{\mb{u}}) \propto
\tr \left(\left[\left(
\nabla_{\mb{u}} \mb{c}^{\tran}(\hat{\mb{u}})\right) \left(
\nabla_{\mb{u}} \mb{c}^{\tran}(\hat{\mb{u}})\right)^{\herm} \right]^{-1} \right) \label{eq:accuracy_approx}.
\end{equation}

In other words, if two similar estimations of $\hat{\mb{u}}$ result in closely related \gls{ris} configurations, the derived position information tends to be less precise. On the other hand, if these estimations lead to significantly different \gls{ris} configurations, the accuracy of the position information improves.

Importantly, to estimate the accuracy, $\mb{c}(\cdot)$ must be known. However, its availability implies knowledge of the \gls{ris} optimization process run at the controller, which might not be available.
However, a possibility is to approximate $\mb{c}(\cdot)$ through a secondary \gls{dnn}.
Let this alternative \gls{dnn} retrieve input parameters from the localization output and reconstruct the corresponding \gls{ris} configuration. The weights of this \gls{nn}, once trained, are approximating $\mb{c}(\cdot)$, and thus, providing an effective and practical approach to compute Eq.~\eqref{eq:accuracy_approx}, even when the \gls{ris} optimization process is unknown.

It is worth underlying that, other than the generation of training data for the \glspl{dnn} involved, \emph{our proposal is independent of the method used to determine the optimal configuration}, with a single condition: the \gls{ris} configurations must depend in a sufficient degree on the \gls{ue} position $\mb{u}$. This condition can be confirmed by evaluating the \gls{fi} that the configurations carry about $\mb{u}$, as explained in Section~\ref{sec:fisher_information}.

Depending on the system setup, the training of \name{} components can be performed in different ways, as follows: \emph{Independently}, if the information of the deployment is available and by exploiting Eq.~\eqref{eq:coherent_solution} to generate a training dataset, or \emph{collaboratively}, in cooperation with the communication infrastructure by retrieving localization data from existing sources. Moreover, if the agent's error distribution is independent of the infrastructure's error, a continuous training process can refine position estimations until the desired accuracy is achieved.

\subsection{Complexity Analysis}

\name{} components are built around \gls{dnn}.
For the complexity analysis, we consider a fully connected \gls{dnn} with $L_0 = N$ input neurons, $L_{\ssub{K}} = D$ outputs, and $K-1$ hidden layers with $L_k$ neurons each. The computational cost of inference directly affects hardware resource usage and is proportional to the number of operations involved.

From the \gls{dnn} topology, the number of multiplications is  $\sum_{k=0}^{K-1} L_k L_{k+1}$, the number of sum operations is $\sum_{k=0}^{K-1} L_k L_{k+1}$, and 
the activation function is computed $\sum_{k=1}^{K} L_k$ times, which reduces to $\sum_{k=1}^{K-1} L_k$ if the identity activation function is chosen for the output layer.
Let us consider hidden layers comprising $L_{\ssub{C}}$ neurons each, yielding to a total of $L_0 L_{\ssub{C}}+(K-2)L_{\ssub{C}}^2+L_{\ssub{K}} L_{\ssub{C}}$ multiplications, $(L_0+L_{\ssub{K}})L_{\ssub{C}}+(K-2)L_{\ssub{C}}^2 $ sums, and $(K-1) L_{\ssub{C}} +L_0$ activation function evaluations.
As a result, the inference computational complexity grows as $\mathcal{O}(L_{\ssub{C}}^2)$, and as $\mathcal{O}(L_0)$. This is empirically confirmed by the experiments described in Section~\ref{sec:embedded_testbed}.
Lastly, it is worth mentioning that the number of cycles required to perform a float number operation is variable in some architectures, which introduces variability in the total time needed for complete inference.

\subsection{FI-guided Complexity Reduction}
\label{sec:complexity_reduction_approach}
Given the typically large number of \gls{ris} elements~\cite{zappone2020optimal}, processing all the inputs requires significant computational effort. This increased complexity could make it challenging to use \name{} alongside real \gls{ris} hardware, which is generally designed to be low-complex.
To address this challenge, we propose a methodology to optimize the implementation of the \gls{dnn} by reducing its computational complexity while preserving high performance. This approach involves strategically reducing both the input features and, consequently, the \gls{dnn} size, guided by the relevance of each input feature to the localization process.

To assess the feature relevance, we compute the \gls{fi} associated with the configuration of each \gls{ris} element according to \eqref{eq:info_per_element}. This metric is averaged over all potential positions of the \gls{ue} within the area of interest, providing a comprehensive measure of the information carried by each input feature. Based on this analysis, we select a subset of \gls{ris} configuration elements to be processed, prioritizing those with the highest \gls{fi} values. The input layer is then resized to match the reduced feature set.
This approach minimizes the computational complexity while retaining the majority of the informative content and preserving the estimation accuracy. In Section~\ref{sec:complexity_reduction_performance}, we demonstrate the effectiveness of this approach.

\section{Numerical Results and Measurements}
\label{sec:results}

\begin{figure}[t]
        \center
        \includegraphics[width=.47\textwidth, trim = {0cm .4cm 0cm 0cm}]{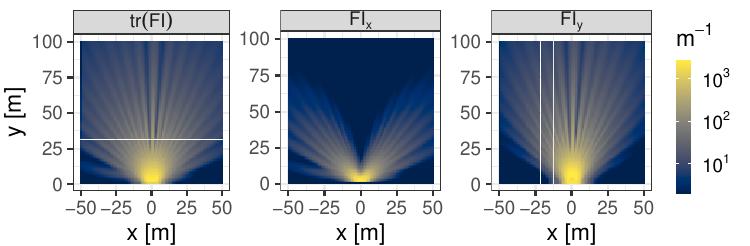}
        \caption{Cumulative Fisher information across all \gls{ris} elements. Computed for a $21\times21$ \gls{ris} with $\sigma_{\uptheta}=\frac{\pi}{6}$ rad, $\mb{p}_{\text{ref}}$ set in the geometric center of the \gls{ris}.}
        \label{fig:FI_complete_RIS}
\end{figure}

To assess the performance of \name{}, we consider a test area of dimensions $100m\times100m$ lying on the $xy$-plane. This area includes a grid of potential \glspl{ue} locations regularly spaced at intervals of $1$ meter in both the $x$ and $y$ directions, spanning $x\in(-50,50)$~m and $y\in(0,100)$~m. The \gls{ue} heights are fixed at $u_z=1.5$ m.
A \gls{ris} is deployed parallel to the $yz$-plane, consisting of $N_y = N_z = 21$ elements, totaling $N = 441$ elements, with geometric center (and $\mb{p}_{\text{ref}}$) located at $\textbf{r} = [0,0,3.5]^{\tran}$ m. The \gls{ris} serves a \gls{bs} with $M=16$ antenna elements, located at $\textbf{b} = [200,1000,50]^{\tran}$ m.
We consider a $1$-bit \gls{ris} phase quantization level ($Q=1$), an operating frequency $f_0 = 6$~GHz, and the path loss exponent of $\beta=2$. The \gls{ris} configuration selection follows the coherent paths approach in Eq.~\eqref{eq:coherent_solution}, while considering the phase error distribution described in Eq.~\eqref{eq:theta_distribution} with $\sigma_{\uptheta} = \{\frac{\pi}{18}, \frac{\pi}{6}, \frac{2\pi}{3}\}$.

\subsection{Feasibility Analysis Through FI and CRB}
\label{sec:feasibility_analysis}

\begin{figure}[t]
        \center
        \includegraphics[width=.47\textwidth, clip = {0cm 0.4cm 0cm 0cm}]{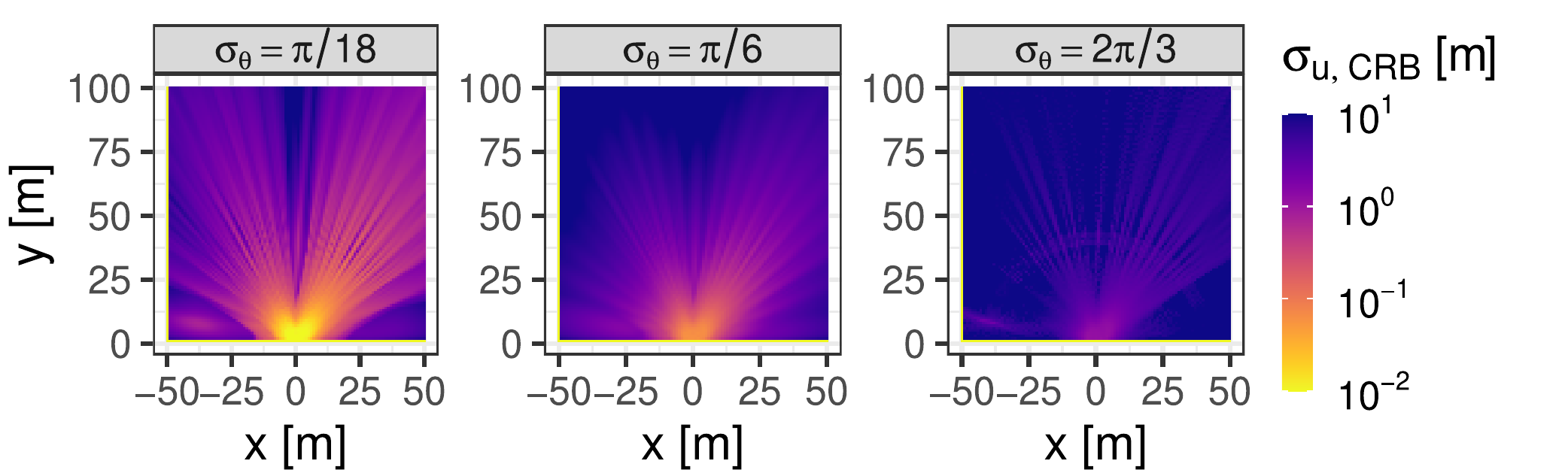}
        \caption{Localization \gls{crb} (Eq.~\eqref{eq:localization_accuracy}) with different noise levels. Computed for a $21\times21$ \gls{ris} with $\mb{p}_{\text{ref}}$ set in the geometric center of the \gls{ris}.} 
        \label{fig:CRB_complete_RIS}
\end{figure}

In the following, we provide a feasibility analysis through \gls{fi} by computing it according to Eq.~\eqref{eq:fisher_info_2} and Eq.~\eqref{eq:fisher_info_binary} for all \gls{ue} locations of the grid, and considering different phase error distributions.

Fig.~\ref{fig:FI_complete_RIS} (left) shows the trace of the \gls{fi} matrix, which serves as a scalar summary of the total amount of information available from the \gls{ris} configuration for different user positions in the service area $(u_x, u_y)$. Results confirm that a non-negligible amount of information on the \gls{ue} position is carried by an optimized \gls{ris} configuration. Nonetheless, the central and the right-hand side plots of Fig.~\ref{fig:FI_complete_RIS} show the amount of information on the $x$ and $y$ directions separately, highlighting that the information on one coordinate might dominate on the other, depending on the \gls{ue} position relative to the \gls{ris}.
Indeed, the trace of \gls{fi} has a limitation: if the information in the $x$ and $y$ directions differs significantly, the trace is dominated by the direction with the highest information, potentially hiding poor performances.

To overcome this limitation, we employ the \gls{crb} as defined in Eq.~\eqref{eq:localization_accuracy}, as its value is dominated by the direction with less information and provides a better view of the achievable accuracy. In other words, the \gls{crb} tells us 
\emph{if and to what extent is it possible to determine the position of the \gls{ue} in all directions.}
The \gls{crb} in the considered service area is shown in Fig.~\ref{fig:CRB_complete_RIS}. We can observe that the user can be located with a sub-m accuracy for a large portion of the area,
which confirms the feasibility of estimating the user location from the sole observation of the \gls{ris} configuration. Fig.~\ref{fig:CRB_complete_RIS} shows the achievable accuracy with different values of $\sigma_{\uptheta}$, the better the \gls{csi} the higher the achievable localization accuracy.

Lastly, we leverage \gls{fi} to measure the information carried by each \gls{ris} element. This provides additional insight on how the information is distributed over the surface and on how to choose the reference element.

\begin{figure}[t]
        \center\vspace{-5mm}
        \includegraphics[width=.376\textwidth, trim = {0cm .4cm 0cm 0.0cm}]{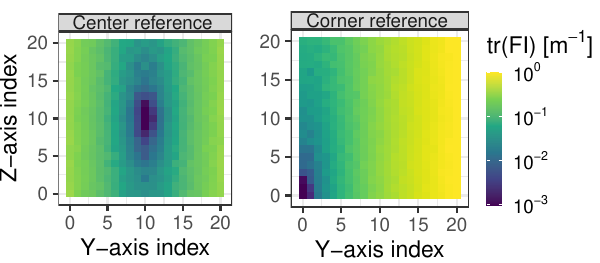}
        \caption{Fisher information from each \gls{ris} element, averaged across all positions in the studied space, following Eq.~\eqref{eq:fisher_info_2}, Eq.~\eqref{eq:info_per_element} and Eq.~\eqref{eq:fisher_info_binary}, for two different $\mb{p}_{\text{ref}}$. }
        \label{fig:FI_on_each_element}
\end{figure}
Results are shown in Fig.~\ref{fig:FI_on_each_element}, where we can observe that the amount of information per element is higher in elements that are far away from the reference one. As a result, choosing the reference element at the corner of the \gls{ris} greatly increases the overall amount of information available for localization purposes. Interestingly, this outcome is in contrast with most of the \gls{ris} configuration solutions in the literature, where the choice of the \gls{ris} reference element is largely irrelevant. Indeed, as long as \gls{ris} elements are configured to constructively interfere at the \gls{ue}, the difference between configurations with different reference elements---while non-zero---is negligible.

\subsection{Localization Accuracy}

To assess \name{} performances, we compute a \gls{ris} configuration for each point of the grid, assuming that the \gls{ue} served by the \gls{ris} is located in that point. The phase noise is set to $\sigma_{\uptheta}=\frac{\pi}{6}$. We obtain a set of $10^4$ pairs of spatial coordinates and configuration that we use for training ($75\%$) and validation purposes ($25\%$).

\begin{figure}[t]
        \center
        \includegraphics[width=.47\textwidth, trim = {0cm .75cm 0cm 0cm}]{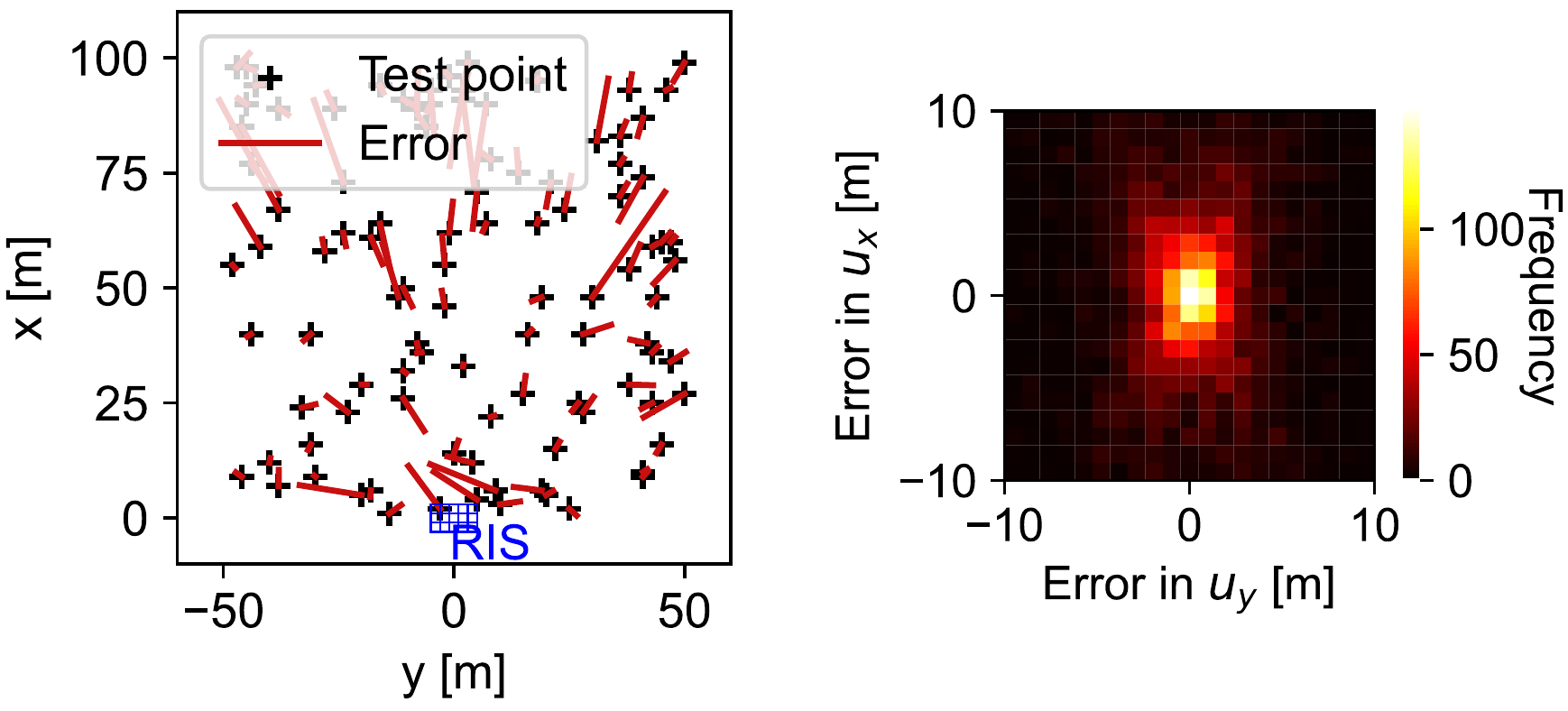}
        \caption{Location estimation performance of \name{} in the considered area.}
\label{fig:NN_estimation} 
\end{figure}

\begin{figure}[t]
        \center
        \includegraphics[width=.47\textwidth, trim = {0cm .75cm 00cm 01cm}]{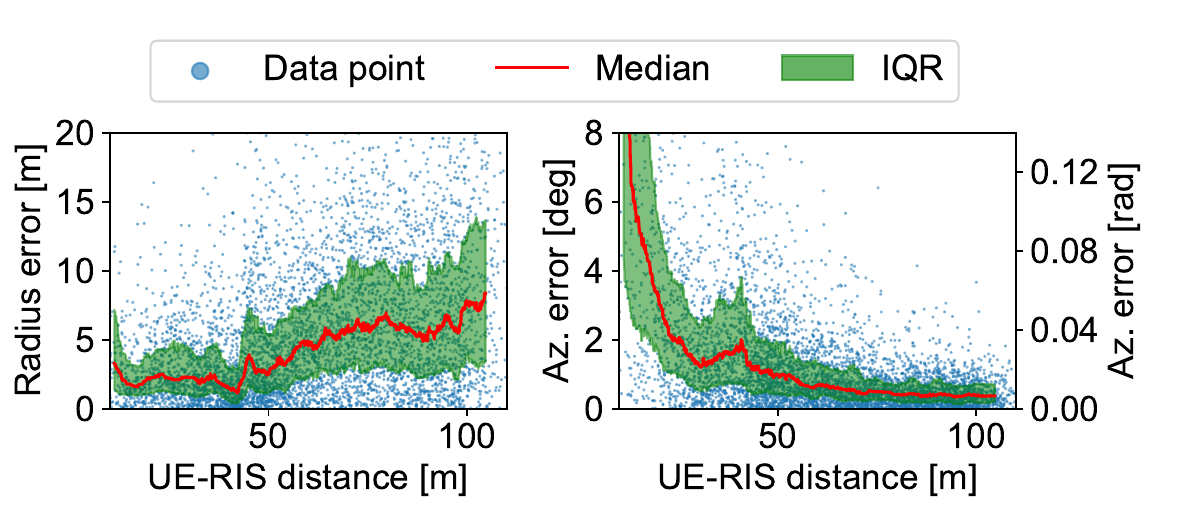}
        \caption{Prediction errors of \name{} translated to polar coordinates w.r.t. the distance \gls{ris}-user, error median and \gls{iqr}.}
\label{fig:NN_estimation_polar} 
\end{figure}
We employ a \gls{dnn} comprising $441$ input neurons, i.e., we process the entire \gls{ris} configuration vector, followed by $2$ fully connected hidden layers of $100$ neurons each, and a $2$ neurons output layer providing the $x$ and $y$ coordinates of the estimation, for a total of $5.4\cdot10^3$ parameters to train.
We determine \name{} performance by running the localization inference on the testing set and comparing the estimation with the true values, i.e., $\|\mb{u}- \hat{\mb{u}}\|$. The average error across the testing set is $5.04$m, for a $5\%$ error in our $100$m scenario. More details are available in Fig.~\ref{fig:NN_estimation}, which shows $100$ randomly selected test points and the corresponding prediction error.
We can observe that the errors exhibit randomized directions in the proximity of the \gls{ris} and a clear radial pattern in farther locations. This behavior reflects the ability of our strategy to accurately determine angularly the \gls{ue} position, and the weak ability to infer its distance when approaching far field conditions. In the right-hand side of Fig.~\ref{fig:NN_estimation} we can see a $2$D histogram of the estimation error over the complete training set.

To further expose the behavior of radial and angular errors against the \gls{ris}-\gls{ue} distance, we show them in polar coordinates with respect to the \gls{ris} in Fig.~\ref{fig:NN_estimation_polar}.
From this, we can see the errors for radius (distance) and azimuth (angle). Both graphs include the original data of all test samples, the median, and the \acrfull{iqr} obtained within a moving window of $100$ samples.
It is noticeable that the radial component of the prediction error increases steadily with distance, while the error in the azimuth does not suffer from this effect. Instead, its accuracy increases at higher distances.

\begin{figure}[t]
        \center
        \includegraphics[width=.8\linewidth, trim = {0cm 0.5cm 0cm 0.0cm}]{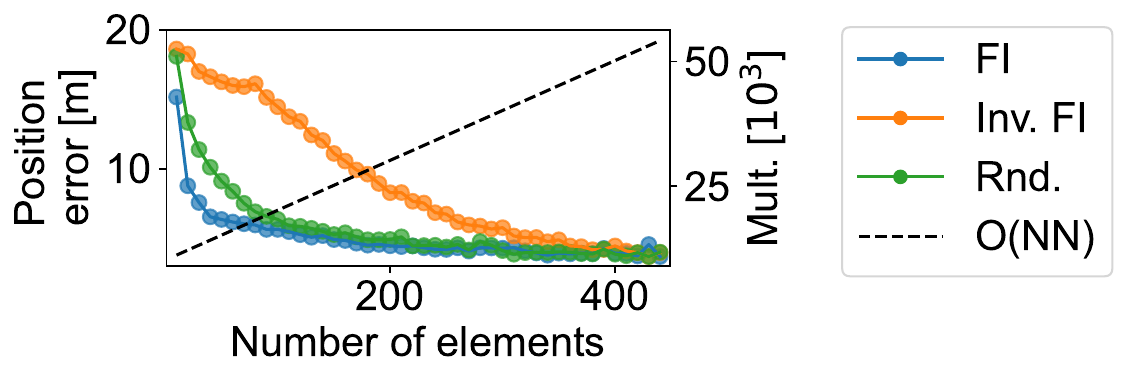}
        \caption{Average \name{} estimation error (left axis) and computational complexity (right axis) against the number of \gls{ris} elements, following different selection strategies.}
        \label{fig:FI_element_selection} 
\end{figure}

\begin{figure}[t]
        \center
        \includegraphics[width=.47\textwidth, trim = {0cm .4cm 0cm 0.0cm}]{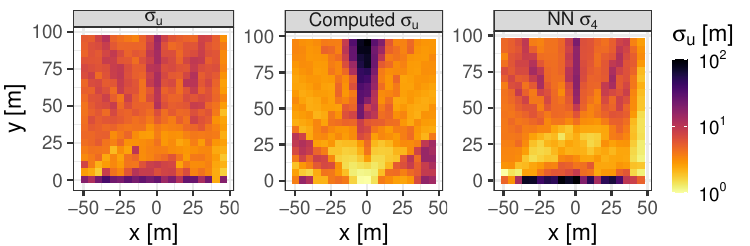}
        \caption{True localization error in comparison with the prediction of \name{} accuracy estimation module.}
        \label{fig:NN2_error_log}
\end{figure}

We now evaluate the ability of \name{} to infer the accuracy of the location estimation. To do so, we compare two methods: one based on the configuration spatial gradient as per Eq.~\eqref{eq:accuracy_approx}, and another based on a second \gls{nn} specifically trained for this purpose.
For the gradient-based approach, we assume knowledge on the \gls{ris} configuration method at \name{}, and we evaluate Eq.~\eqref{eq:accuracy_approx} in the grid points. The result is multiplied by a constant value that is selected in such a way that the prediction is unbiased.
Hence, the average predicted error matches the real one.
Differently, for the \gls{nn} approach, we employ a \gls{dnn} that processes the whole \gls{ris} configuration, comprising two hidden layers of $100$ neurons and producing the magnitude of the error as output. 
The dataset for training and testing this \gls{dnn} is created from the test set of the first \gls{dnn} by pairing the magnitude of its inference error with the corresponding \gls{ris} configuration.
To have enough data to train both the localization and the accuracy prediction \glspl{dnn}, we set $\Delta = 0.25$~m, generating $1.6\cdot10^5$ data instances.
In particular, the obtained data set is divided into training ($43\%$) and testing ($57\%$) subsets for the localization \gls{dnn}, and the testing subset is further divided for training ($75\%$) and testing ($25\%$) the \gls{dnn} estimating the accuracy.
Fig.~\ref{fig:NN2_error_log} shows the predicted accuracy against the actual one with the different estimation methods and confirms the reliability of \name{} in the accuracy prediction task.

\subsection{Comparison with traditional \gls{aoa} methodology}
\label{sec:soa_aoa}
\begin{figure}[t]
    \centering
    \begin{minipage}[c]{0.23\textwidth}
        \centering
        \includegraphics[width=\textwidth, trim={0cm .0cm 0cm 0cm}]{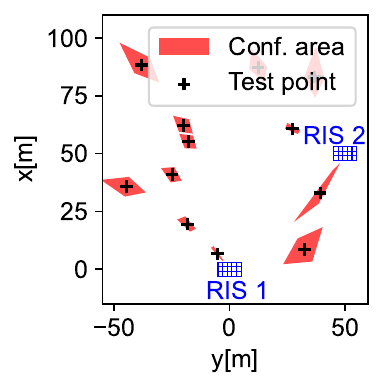}
    \end{minipage}
    \hfill
    \begin{minipage}[c]{0.23\textwidth}
        \centering
        \includegraphics[clip, width=\textwidth, trim={0cm .0cm 0cm 0cm}]{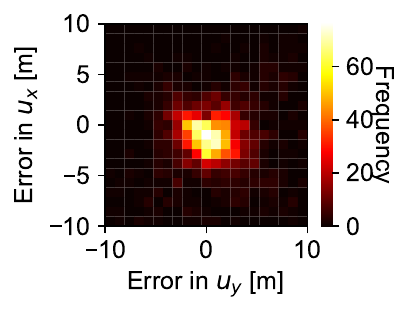}  
    \end{minipage}
    \caption{Location estimation performance of a traditional \gls{aoa} schema.}
    \label{fig:NN_estimation_trad} 
\end{figure}

We consider a traditional \gls{aoa}-based localization schema using \glspl{ris}~\cite{xu2008aoa}. In particular, we employ two \glspl{ris} devices in the area, one in $(0,0)$ and one in $(50,50)$, that locate the user by scanning the area with directional beams until connecting the user and intercept their beams locate it. Beams have a horizontal \gls{hpbw} equal to $\frac{0.886 \lambda}{N_y d \cos{\phi}}$~\cite{balanis2015antenna, mailloux2017phased}, $d$ is the inter-element distance and $\phi$ is the beam angle with respect to the normal direction to the surface, which we assume directed to the user location. We take the \gls{hpbw} as the angular information uncertainty.
In Fig.~\ref{fig:NN_estimation_trad}, we plot the confidence areas of a few users (left-hand side) and a histogram of the errors over the complete dataset (right-hand side).
The obtained accuracy is slightly better than the one of \name{} (Fig.~\ref{fig:NN_estimation}). However, it requires the simultaneous employment of two coordinated devices to perform localization.

\subsection{Operation in \Gls{nlos} conditions}
\label{sec:nlos_operation}
\begin{figure}[t]
    \centering
    \begin{minipage}[c]{0.23\textwidth}
        \centering
        \includegraphics[width=\textwidth, trim={0cm .0cm 0cm 0cm}]{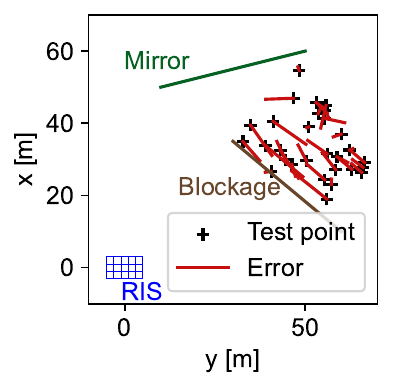}  
    \end{minipage}
    \hfill
    \begin{minipage}[c]{0.23\textwidth}
        \centering
        \includegraphics[clip, width=\textwidth, trim={0cm .0cm 0cm 0cm}]{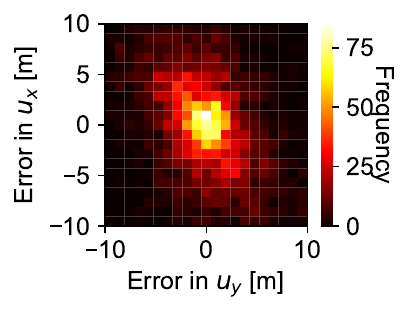}
    \end{minipage}
    \hfill
    \begin{minipage}[c]{0.23\textwidth}
        \centering
        \includegraphics[clip, width=\textwidth, trim={0cm .0cm 0cm 0cm}]{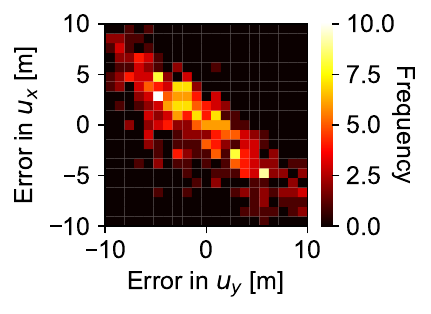} 
    \end{minipage}
    \caption{Performance of \name{} in NLoS. Top-left: \gls{nlos} estimations and errors. Top-right: test set error histogram. Bottom: \gls{nlos} error histogram.}
    \label{fig:NN_estimation_NLoS} 
\end{figure}

After evaluating the performance of \name{} in \gls{los} condition between the \gls{ris} and the test points, we now consider its operation in \gls{nlos}.\footnote{While the most commonly proposed use case of \glspl{ris} is to operate in \gls{los} regime to improve radio links at high frequencies, there is no fundamental limitation of \name{} in this regard.\label{foot:nlos}}
To test and benchmark our proposal in \gls{nlos} conditions, we modify the scenario to include reflections. In particular, in the $100m\times100m$ area, we introduce two additional elements: a mirror and a blockage, as depicted in Fig.~\ref{fig:NN_estimation_NLoS}. The blockage completely absorbs the impinging signal, while the mirror specularly reflects it.
We collect test points similarly as in the previous simulations, but discarding those that, from the point of view of the \gls{ris}, fall behind the mirror, and those that are blocked by the obstacle without a viable reflection to reach the \gls{ris}.
In particular, a viable reflection exists if, when considering the specular image of the test point with respect to the plane of the mirror, the segment delimited by the specular image and the \gls{ris} position intersects with the mirror. Test points located behind the blockage but reachable through reflection by the \gls{ris} are kept.

In the upper left of Fig.~\ref{fig:NN_estimation_NLoS} we represent a subset of those test points, the mirror, and the blockage. For clarity, other test points are not in the picture. Nonetheless, any test point in the area not shadowed by the blockage or the mirror is included in the training set, as in Fig.~\ref{fig:NN_estimation}.
In the upper right of Fig.~\ref{fig:NN_estimation_NLoS} we reproduce the $2$D histogram of the estimation errors of the complete test set, and in the bottom of Fig.~\ref{fig:NN_estimation_NLoS} we can see the same graphic but taking only the \gls{nlos} points of the same set. 

The localization error related to \gls{nlos} scenario is slightly above the average error considering the complete set (as can be seen by comparing Fig.~\ref{fig:NN_estimation_NLoS} with Fig.~\ref{fig:NN_estimation}). Moreover, the error pattern of the \gls{nlos} points follows a preferred direction in this case, specifically the radial direction in the specular image of the points.

Overall, the performance of \name{} in mixed \gls{los} and \gls{nlos} conditions is remarkably good, being able to finely distinguish reflected and non-reflected points, even when they exhibit very similar \gls{ris} configurations.

\subsection{Complexity Reduction Performance Assessment}
\label{sec:complexity_reduction_performance}
With the objective to run \name{} in the embedded platform while consuming a minimal amount of energy, we reduce the initial \gls{nn} design to minimize the inference computational complexity. While reducing the hidden layers dimension is trivial, selecting among the vast amount of input variables poses a different problem, advanced in ~Section~\ref{sec:feasibility_analysis}.
In this case we employ the \gls{fi} to reduce the number of input parameters to be processed.
In particular, we consider the \gls{fi} of each element, averaged over the possible \gls{ue} positions in the area (Fig.~\ref{fig:FI_on_each_element}),
 and select a subset of them based on the amount of information carried, we train a \gls{nn} and individually evaluate its average error.
In Fig.~\ref{fig:FI_element_selection}, we show the resulting localization accuracy obtained with different \gls{ris} element selection strategies together with the resulting complexity of the inference process.
In particular, the blue series incorporates the first elements sorted by descending \gls{fi}, the orange series incorporates elements in random order and the green series incorporates the elements by ascending \gls{fi}.
The \gls{fi}-guided strategy with higher \gls{fi} elements selection outperforms both the random selection and the \gls{fi}-guided strategy with lowest \gls{fi} values. In particular, it is observable how a high fraction of the best performance can be achieved with the observation of a small portion of the total elements.

For the \name{} implementation in the embedded platform, we select an input vector of $81$ values and resize the hidden layers to $4$ neurons each, leading to a total of $358$ \gls{dnn} parameters to tune, which is approximately the $0.6$\% of the previous amount.
Despite this abysmal decrease in the required training effort and computational complexity of the individual inferences, the average estimation error only grows up to $11.40$m. 

Finally, we evaluate this reduced \gls{dnn} in indoor-like settings by considering a smaller area of $10 \times 10$~m, with a smaller separation between points of $\Delta = 0.1$~m. We place a $N_y = N_z = 9$ \gls{ris} and proceed as with the previous experiment, yielding a performance of $0.70$~m average localization error. This reduced \gls{nn} is implemented and tested in the testbed we describe in the next section.

\section{\name{} Embedded Prototype}
\label{sec:embedded_testbed}

\begin{figure*}
        \center
        \includegraphics[width=\linewidth, trim = {-0.05cm 0cm 0cm 0cm}]{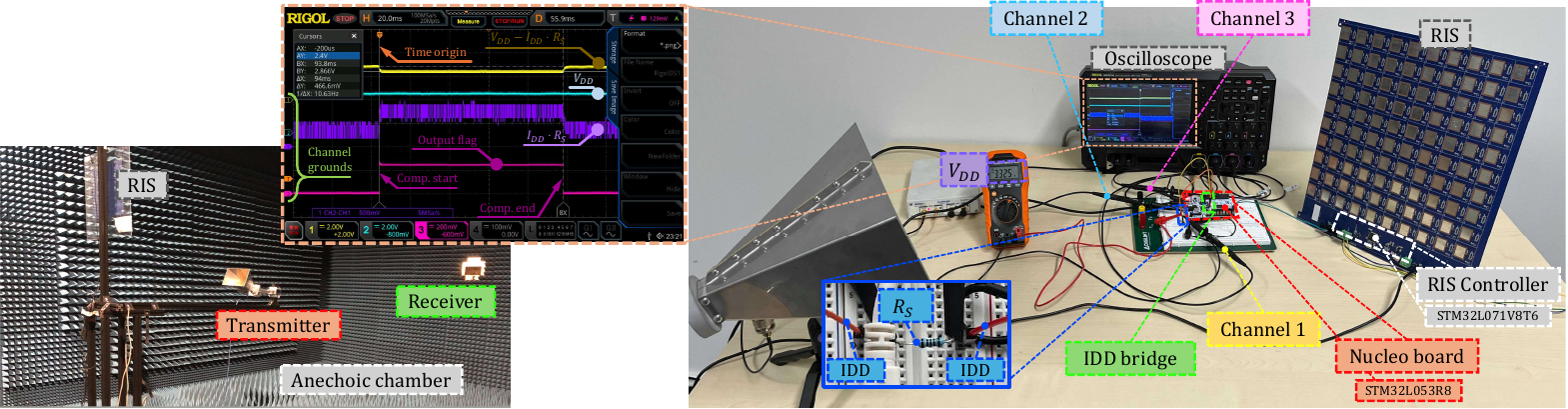}
        \caption{Testbed setups with an oscilloscope screen while performing the position inference for a \gls{ris} configuration.}
        \label{fig:testbed_photo} 
\end{figure*}

\name{} localizes the \gls{ue} by the direct processing of the \gls{ris} configuration. Hence, a practical and convenient approach is to deploy it at the \gls{ris} (or at the \gls{ris} controller).
Such settings are only viable if \name{} can cope with the stringent computational and power consumption limitations of commercial \glspl{ris} hardware~\cite{NEC-ris}.
With this in mind, we prove the practical feasibility of \name{} by deploying into a lightweight, low-power embedded platform, which is in line with the commercial \gls{ris} hardware characteristics.
To this aim, we select the STM32L053R8 Arm\textregistered\space\gls{mcu}, from STMicroelectronics N.V., which is a low-power Arm\textregistered-based Cortex\textregistered-M0+ \gls{mcu}, equipped with $32$-bit wide arithmetic logic unit and data bus, $64$ KB of non-volatile flash memory, and $8$ KB of random access memory~\cite{cortexm0programmingmanual, datasheet053mcu}.
Additionally, it is a viable option for low-cost devices thanks to its low price and minimal external hardware requirements.

\begin{algorithm}[t!]
\scriptsize
    \caption{Pseudocode of the \gls{nn} generation for ARM testbed firmware
\label{alg:training}} 
    \begin{algorithmic}[1]
        \State Set $\mb{b}$, $\mb{r}$, $f_0$, $\textbf{p}_{n}$ $\forall n \in [1,N]$, and $\mb{p}_\text{ref}$ \Comment{\textit{\color{blue}{\scriptsize{Scenario definition}}}}
        \State Generate $\mb{u}_{i,j}$ values \Comment{\textit{\color{blue}{\scriptsize{Create \gls{ue} positions grid}}}}
        \State Set RNG, $\sigma_{\theta}$, and $N_{\ssub{E}}$ \Comment{\textit{\color{blue}{\scriptsize{Random number generator, noise and number of examples}}}}
        \For {all points $\mb{u}_{i,j}$ } \Comment{\textit{\color{blue}{\scriptsize{Iterate over grid points}}}}
            \State Compute $\mb{H}_{\ssub{RB}}, \mb{h}_{\ssub{RU}}, \mb{h}_{\ssub{D}}$ \Comment{\textit{\color{blue}{\scriptsize{Actual channel conditions}}}}
            \For {$\forall e \in [1,N_{\ssub{E}}]$} \Comment{\textit{\color{blue}{\scriptsize{Iterate over number of examples}}}}
            \State Generate noise and compute $\hat{\mb{H}}_{\ssub{RB}}, \hat{\mb{h}}_{\ssub{RU}}, \hat{\mb{h}}_{\ssub{D}}$ \Comment{\textit{\color{blue}{\scriptsize{Noisy channel conditions}}}}
            \State Compute $\bar{\boldsymbol{\uptheta}}_{e,i,j} = \mb{g}(\hat{\mb{H}}_{\ssub{RB}}, \hat{\mb{h}}_{\ssub{RU}}, \hat{\mb{h}}_{\ssub{D}})$ (Eq.~\eqref{eq:coherent_solution}) \Comment{\textit{\color{blue}{\scriptsize{\gls{ris} configuration}}}}
        \EndFor
        \EndFor
        \State Randomly sort and split $(\bar{\uptheta}_{e,i,j}, u_{i,j})$ pairs \Comment{\textit{\color{blue}{\scriptsize{Train and test set creation}}}}
        \State Define \gls{nn} topology and activation functions  \Comment{\textit{\color{blue}{\scriptsize{\gls{nn} initialization}}}}
        \State Train \gls{nn} until convergence using train subset \Comment{\textit{\color{blue}{\scriptsize{Training phase}}}}
        \State Measure inference accuracy using test subset \Comment{\textit{\color{blue}{\scriptsize{Testing phase}}}}
        \State Extract weights and biases of the \gls{nn} \Comment{\textit{\color{blue}{\scriptsize{End}}}}
    \end{algorithmic} 
\end{algorithm}

\begin{table}[h!]
\setlength{\abovecaptionskip}{0pt}
\setlength{\belowcaptionskip}{0pt}
\caption{Testbed setup parameters}
\label{tab:parameters}
\centering
\resizebox{\linewidth}{!}{
\begin{tabular}{cc|cc}
\textbf{Parameter} & \textbf{Value} & \textbf{Parameter} & \textbf{Value}\\  
\hline
\rowcolor[HTML]{EFEFEF}
MCU  & STM32L053R8 & $V_\text{DD}$  & $3.31$ V    \\
CPU Clock (HCLK)  & $2.096$ MHz &  NN Input size & $81$\\
\rowcolor[HTML]{EFEFEF}
NN Hidden layers size & $(4,4)$ &  NN Output size & 2   \\
Number of parameters & $358$ & $R_{\ssub{S}}$  & $0.994$ k$\Omega$ \\
\hline
\end{tabular}%
}
\end{table}

Fig.~\ref{fig:testbed_photo} shows the experimental setup we use, while Table~\ref{tab:parameters} details the setup parameters. 
We employ the NUCLEO-L053R8 development board~\cite{usermanualnucleo},
which allows to run the \gls{mcu} with all the needed circuitry, conveniently access all its terminals and includes a communications and programmer circuit section called ST-Link/V2 ~\cite{boardschematicnucleo}.
The \gls{mcu} can receive commands and print the result of inference computations through the \gls{usart}
peripheral~\cite{datasheet053mcu},
as described in Algorithm~\ref{alg:firmware}. The input and output pins used by this interface are connected through the ST-Link/V2 part of the Nucleo board. The board can then connect with a PC through a \gls{usb} port, presenting itself as a virtual COM port.
We use the external PC for the training process of the embedded \gls{nn} as described in Algorithm~\ref{alg:training}, to create, compile, and load the firmware into the \gls{mcu} (including the \gls{nn} created with the weights obtained in the previous training stage), to send commands and configurations to the board, and to receive the results and feedback back from the board.
We use a KLEIN TOOLS MM400 multimeter and a RIGOL MSO5104 oscilloscope for measurements.

To measure execution times and estimate power consumption, we monitor the current drawn by the microcontroller during inference. 
The connection labeled as \textit{IDD} is the only bridge between the \gls{mcu} and its power supply ${V}_\text{DD}$.
We remove the IDD bridge, substitute it with a shunt resistor $R_{\ssub{S}}$ of $1.00$ $\text{k}\Omega$ and $1\%$ tolerance, and monitor the voltage at both $R_{\ssub{S}}$ ends with the oscilloscope.
Given the $R_{\ssub{S}}$ value, we can easily infer the absorbed current from the measured voltage, effectively serving as an ammeter. Alternatively, to obtain a precise measurement of sustained currents, we substitute the IDD jumper for a dedicated ammeter.

Finally, we employ the pin PC8 in the CN10 connector of the board as an external flag, to precisely mark different events in the code, such as the beginning or end of an inference or a low-power sleep mode.
The PC8 pin is controlled by our code and serves as an external flag pin in the firmware. We monitor its status through the \textit{channel 3} of the oscilloscope.

\subsection{Energy Consumption}
\label{sec:energy_consumption}
\begin{algorithm}[t!]
\scriptsize
\caption{Pseudocode of the ARM testbed firmware}
\label{alg:firmware}
\begin{algorithmic}[1]
    \State Reset peripherals, initialize flash interface
    \State Initialize ARM SysTick, System Clock, GPIOs, USART, and timers
    \State Instantiate \gls{nn}, load weights and biases, and start timer
    \For {Program execution}
        \State Check USART buffer
        \If {Configuration received}
            \State Record configuration, record timer (start)
            \State Raise external flag pin
            \State Perform \gls{nn} inference
            \State Lower external flag pin
            \State Record timer value (end)
            \State Send back inference result, timer difference, processed configuration
        \EndIf
        \If {Sleep command received}
        \State Enter sleep mode until interrupt
        \EndIf
    \EndFor
\end{algorithmic}
\end{algorithm}

To analyse the computational resources and the energy used by \name{} we intercept the power supply and measure an output pin of the \gls{mcu} as detailed in Section~\ref{sec:embedded_testbed}. In Fig.~\ref{fig:testbed_photo}, we can see a measurement under such a setup: The oscilloscope has been configured to capture a single event, using the external flag as a trigger on raised edges. That way the time zero, marked by an irregular orange pentagon with an uppercase \textit{T} is located when channel $3$ (in pink, labeled \textit{3}) is raised.
While the large shunt resistor and the channels $1$ and $2$ that measure the voltage at its ends allow us to have a rough estimation of the power consumption, and to see the consumption trends and changes, the voltage drop is too small in comparison with the sensitivity of this oscilloscope for accurately measuring the current.
Hence, we substitute the shunt resistor with an ammeter. Moreover, we run an infinite computation loop version of Algorithm~\ref{alg:firmware} to continuously measure the current flow.
In this regime, the oscilloscope is used to confirm that the time out of the computation state is negligible, and the computation ${I}_\text{DD}$ is directly measured. The results of our measurements can be seen in Table~\ref{tab:measurements}.

\begin{table}[h!]
\setlength{\abovecaptionskip}{0pt}
\setlength{\belowcaptionskip}{0pt}
\caption{Energy consumption and execution time}
\label{tab:measurements}
\centering
\resizebox{\linewidth}{!}{
\begin{tabular}{cc|cc}
\textbf{Parameter} & \textbf{Value} & \textbf{Parameter} & \textbf{Value}\\  
\hline
\rowcolor[HTML]{EFEFEF}
\gls{ris} power consumption  & $63.7$ mW &  ${I}_\text{DD}$ waiting & $501$ µA \\
${I}_\text{DD}$ computing & $526$ µA &  ${I}_\text{DD}$ sleep & $229$ µA \\
\rowcolor[HTML]{EFEFEF}
Min computing time & $84.15$ ms &  Max computing time & $102.68$ ms \\
Computing power &  $1.74$ mW &  Max computing energy & $179$ µJ \\
\hline
\end{tabular}%
}
\end{table}

The measured absorbed power during inference is $1.74$ mW, corresponding to the $2.7$\% of the \gls{ris} power consumption. For the sake of comparison, the total energy stored in a pair of AA batteries (holding each $2000$ mAh of charge at $1.5$ V) could run the device for more than a year while performing more than $3\cdot10^5$ position estimations per day, for a staggering total of more than $120$ million position estimations. Moreover, the absorbed power is well below the one which is collectible through \gls{ris}-oriented energy harvesting means~\cite{albanese2023ares}. Additionally, the \gls{mcu} clock frequency of $2.096$ MHz is supported also with $V_{\text{DD}} = 2$ V~\cite{datasheet053mcu}, further prolonging the operational lifetime of the device.
It is also worth noting that a large portion of the \gls{mcu} power consumption is due not to the computation of the position estimations, but to the active peripherals we employ to measure execution times and to control the device. This can be observed in the large ${I}_\text{DD}$ still present in sleep mode, when the CPU of the microcontroller is stopped.

The worst measured inference time with \gls{mcu} running at $2.096$ MHz is $102.68$ ms. However, the \gls{mcu} can be set to run up to $32$ MHz, resulting in a reduced computing time of approximately $1/15$ of the original amount. Despite the higher clock rate, we measure a similar energy consumption per inference, which suggests that in practical deployments, the overall energy consumption is primarily determined by the frequency with which the location estimation is carried out. As a result, the frequency of localization estimation could be adapted to specific use cases and application needs ensuring energy-efficient operation across various scenarios. Importantly, the measured inference time is comparable with the reconfiguration time of current \gls{ris} prototypes, which is in the order of tens of ms~\cite{rossanese2022designing, alexandropoulos2023ris}. This ensures that with a proper selection of the \gls{mcu} clock, \name{} can continuously estimate positions without sacrificing timeliness.

\begin{figure}[t]
        \center
        \includegraphics[width=.47\textwidth, trim = {0cm .75cm 0cm 0cm}]{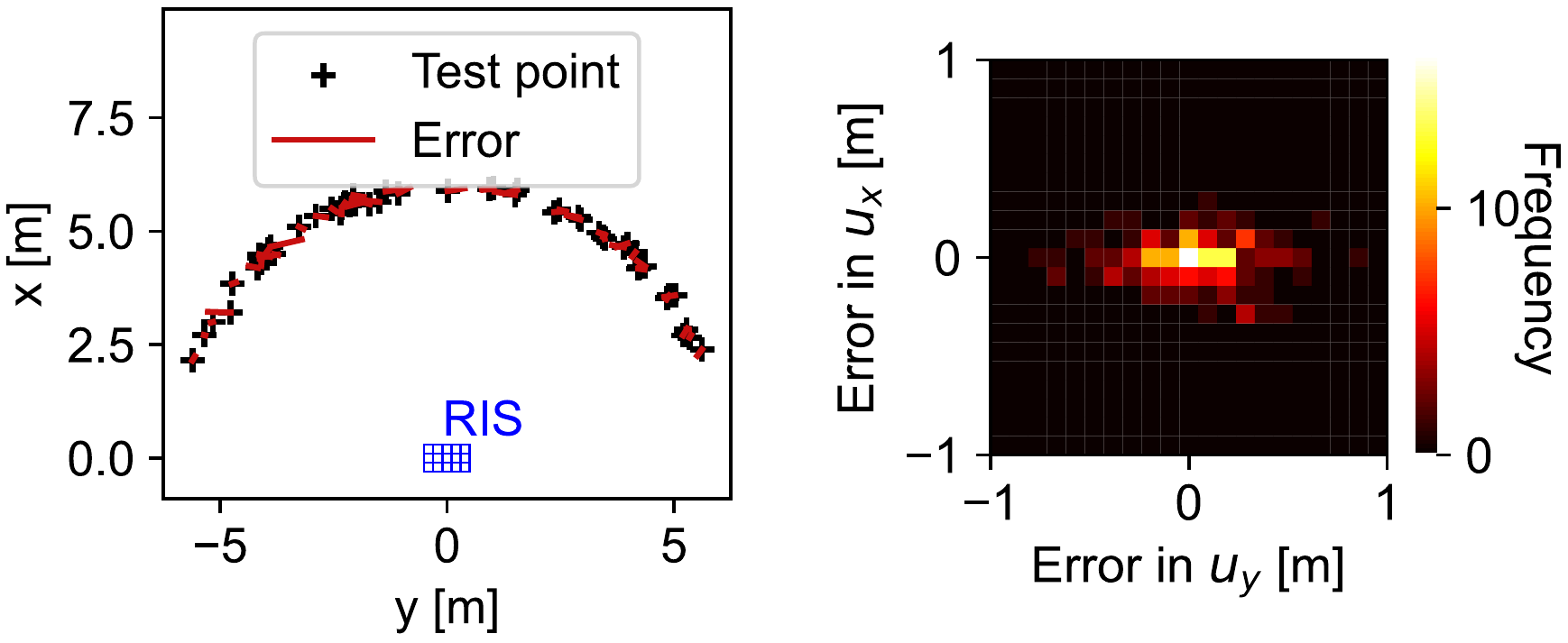}
        \caption{Predictions errors of \name{} operating with commercial \gls{ris} configuration data.}
\label{fig:NN_ris_anechoic}
\end{figure}

\subsection{Opportunistic ISAC Proof-of-concept}
\label{sec:poc}
We set up our experimental environment by means of a commercial RIS device~\cite{NEC-ris}, which is dynamically configurable. In particular, we implement a complete chain including a configuring agent that sends selected \gls{ris} configurations to our \name{} device, completing the diagram in Fig.~\ref{fig:coloris_operations}. While we have full control of the \gls{ris}, we implement the controller as an independent device to better embody the foreseeable commercial setups we focus in this work: a standalone \gls{ris} that either configures itself with or without external feedback or is configured by a centralised controller of the communication network. We assume that we cannot alter that and proceed to use information in a minimally invasive way. We first carry out a measurement campaign in an anechoic chamber (see Fig.~\ref{fig:testbed_photo}) keeping track of the received power gain obtained when signals from a fixed spot reflect in the \gls{ris}. Then, we fix the transmitter position, define a set of test points in front of the \gls{ris}, and for each point, we measure the gain obtained from the $1860$ configurations available in the \gls{ris} codebook. Thus, we select for each test point the configuration providing the highest gain. 
We use the set of possible positions of the receiver and their corresponding configurations as our base dataset. 

Due to limitations in the measurement process, our dataset captures only a segment (arc, see Fig.~\ref{fig:NN_ris_anechoic}) of constant radius ($\rho = 6$m), featuring $47$ points spread across a range from $-70$ to $+70$ degrees. To enhance this dataset for machine learning purposes, we have adopted a noise augmentation strategy, previously validated in other \gls{ml} scenarios \cite{brown2003use, fan2021classification}.

To enhance the dataset's utility for machine learning, we start by normalizing the original data with a standard deviation $\sigma=1$.
Next, we augment this normalized data by creating additional copies, each infused with Gaussian noise of $\sigma_N = 0.05$. This results in $15$ synthetic versions of the original dataset, effectively increasing the diversity and volume of data available for training. The augmented dataset is divided into a training set, for teaching the neural network to predict user locations from \gls{ris} configurations, and a test set, for evaluating its predictive accuracy. This process ensures the network is trained and then tested on unseen data, with its performance assessed by comparing its location predictions to actual data points. The results are detailed in Fig.~\ref{fig:NN_ris_anechoic}. 
The average error in the testing set is $0.41$m, corresponding to the $11$\% of the range of the test points used in the scenario.

We hence validate our proposal, showing \name{} cooperating with a running \gls{ris} to accurately locate users without interfering with the communication architecture and given only access to the configuration of the \gls{ris}. 

\section{Related Work}
\label{sec:related}

In recent years, \glspl{ris} have aroused significant interest in industry and academia for their capacity to control the propagation environment at will~\cite{renzo2019smart,xu2023reconfiguring}.
This unique property puts \gls{ris} as a forefront technology for enhancing the performance and service quality of next-generation networks. Its potential has been explored in several applications, ranging from network spectral efficiency improvement~\cite{mursia2020risma} to enhancing service area coverage ~\cite{albanese2022ris} and support for aerial-to-ground communications~\cite{devoti2023taming}. Additionally, \gls{ris} is promising in facilitating first aid communication infrastructure deployment~\cite{mursia2021rise}.

In addition to supporting communication, \gls{ris}-assisted sensing and localization applications are novel and emerging applications~\cite{zhang2020towards}.
Focusing on the localization use case, in~\cite{emenonye2023fundamentals}, authors leverage on the \gls{fi} matrix to investigate the bounds on the localization accuracy achievable with \gls{ris} aided communications.
A two-stage system based on statistical beamforming for the direction of arrival and time of arrival estimation in \gls{ris}-aided networks is proposed in~\cite{albanese2021papir}. In~\cite{zhang2023multi}, \gls{ris} are used to optimize the multipath profile of the channel for improving time of arrival estimation-based localization under \gls{nlos} conditions. In~\cite{zhang2021metalocalization}, \glspl{ris} are employed to improve the performance of reference signal strength-based indoor localization techniques. While in~\cite{keykhosravi2023leveraging}, authors discuss scenarios where localization would be infeasible for conventional non\gls{ris} techniques. A framework for \gls{ris}-assisted localization under near-field conditions is proposed in~\cite{luan2021phase}, which focuses on \gls{ris} configuration design to perform near-field localization in \gls{nlos} condition from the signal reflected to an anchor point. In the near-field regime, in~\cite{rahal2022constrained}, authors propose an \gls{ris} phase profile design optimization framework tailored to the localization performance improvement, in \cite{wang2022location} authors investigate the effect of \gls{ris} in \gls{isac} by jointly quantifying communication and localization performance, finally~\cite{icct2023merouane} proposes a superimposed symbol scheme for \gls{isac} that overrides sensing pilots onto data
symbols using the same time-frequency resources.

However, \gls{soa} \gls{ris}-aided localization and sensing techniques are largely based on already existing methods, wherein active devices (e.g., \gls{bs} and \glspl{ue}) are already performing operations for sensing and localization, while the \gls{ris} is actively controlled to enhance accuracy.
The application of \gls{ris} in radar surveillance under \gls{nlos} conditions is investigated in~\cite{aubry2021ris}.
A \gls{tdoa} localization algorithm specifically designed for \gls{nlos} conditions using \gls{ris} is studied in~\cite{chen2023tdoa}. In~\cite{nuti2023ris}, a joint scheme that combines preamble detection with localization has been developed, which utilizes a single passive \gls{ris} to improve localization both in \gls{los} and \gls{nlos} conditions.
Practical algorithms have been proposed that reduce computational complexity while leveraging large \gls{ris} setups to localize users under \gls{nlos} conditions have been proposed in \cite{dardari2021nlos,dardari2021localization}. The potential of \gls{ris} for improving radio localization and mapping in \gls{nlos} conditions is discussed in~\cite{wymeersch2020radio}, while channel estimation techniques for improved localization under \gls{nlos} conditions have been faced in~\cite{han2022localization}.
While in~\cite{ge2022ris}, \glspl{ris} are applied in the field of spectrum sensing applications, demonstrating noticeable potential in detection performance improvement. 
Moreover, \gls{ris} is seen as a promising technology also in the \gls{isac} realm~\cite{liu2023integrated,wang2022location}, where it is seen as an enabler for adaptive communication and sensing functionalities~\cite{cheng2023nested}, overcoming the communication drawbacks introduced by the \gls{isac} waveform design~\cite{zhong2023joint}, and for achieving ubiquitous sensing and communication capabilities at the network~\cite{zhang2022toward}.

Several works are considering hardware equipping smart devices, such as cameras and compasses, for opportunistic sensing applications~\cite{nemati2017opportunistic}. However, only a few of them rely opportunistically on communication operations to perform sensing. For instance, in~\cite{ding2018robust}, phase changes in the measured \gls{csi} are employed for motion detection. In~\cite{devoti2020pasid}, power measurements of directional devices performed during the beam alignment procedure are used to monitor propagation conditions indoors to detect and locate people. Similarly, in~\cite{devoti2022passive} adopts a comparable approach to measure crowd levels in public areas. Another example is in~\cite{giannetti2023opportunistic}, where the signal quality of satellite microwave links is used to sense rainfalls.

In contrast with the aforementioned \gls{soa}, we propose a novel \gls{isac} technique that leverages the configuration of an \gls{ris} operating for communication enhancement to opportunistically perform localization. Importantly, the proposed approach eliminates the need for specific operations at the \gls{ris} tailored to the localization task. Additionally, to the best of our knowledge, it represents the first attempt to opportunistically extract location information from the \gls{ris} embedded configurations.

\section{Conclusions}

In this paper, we have explored the transformative potential of \acrfull{ris} within the context of sustainable 6G \acrfull{isac} systems. 
Our goal has been addressing the challenge of user localization within a communication network by harnessing the controlled-reflective properties of \gls{ris} devices. We designed and evaluated \name{}, which demonstrated to achieve accurate user localization without relying on more energy-hungry traditional methods such as GPS or the need of deploying additional infrastructure.

Moreover, our Opportunistic \gls{isac} framework leverages non-intrusive localization-agnostic \gls{ris} configurations to position network users through trained learning models for limited systems (embedded, low-power).
We conducted an experimental evaluation campaign that validated the effectiveness of our approach, demonstrating a positioning accuracy in the $5\%$ range for a $100$m $\times$ $100$m scenario. Finally, the low-complexity solution for limited systems was evaluated with off-the-shelf equipment and showcased its practical feasibility in real-world applications while demonstrating an energy consumption low enough to do neural network inferences for a year while powered with a coin battery.

\begin{appendices}
\section{Proof of Gaussianity of $\hat{\mb{\uptheta}}$}
\label{sec:proof_of_gaussian_assumption}
Let us consider a noisy channel sample $h \!= \!u \!+ \!jv \in \mathbb{C}$, with $u$ and $v$ being two statistically independent \glspl{rv} distributed as
$u \!\sim\! \mathcal{N}\left(\mu_u, \sigma^2/2\right)$ and $v \!\sim\! \mathcal{N}\left(\mu_v, \sigma^2/2\right)$.
$h$ is then distributed as a circular Gaussian \gls{rv}, i.e, $h \sim \mathcal{CN}(\mu_{h},\sigma^2)$, with \gls{pdf}
\begin{align}
f_{h}(h) \!\!=\!\! \tfrac{1}{\pi\sigma^2}e^{-\tfrac{|h-\mu_{h}|^2}{\sigma^2}} \!\!=\!\! f_{u,\!v}(u,\!v) \!\!=\!\!  \tfrac{1}{\pi\sigma^2} e^{-\tfrac{(u-\mu_u)^2 + (v-\mu_v)^2}{\sigma^2}},
\end{align}
where $\sigma^2$ and $|\mu_h|^2 = \mu_u^2 + \mu_v^2$ are the noise and signal power, respectively.

Let us introduce two new \glspl{rv}, $r$ and $\theta$, such that
$r = |h| = \sqrt{u^2 + v^2}$ and $\theta = \angle h = \tan^{-1}\left(v/u\right)$. Let us recall also the relationship $u = r \cos \theta$ and $v = r \sin \theta$.
The joint distribution of $r$ and $\theta$ is obtained with
\begin{align}
f_{r, \theta}(r, \theta) = f_{u, v}(g_1^{-1}(r), g_2^{-1}(\theta))|J| \label{eq:joint_distr_1}
\end{align}
where $g_1^{-1}(r) = r \cos \theta$, $g_2^{-1}(\theta) = r \sin \theta$, and $\lvert J\rvert$ is the Jacobian of the transformation, which is
\begin{align}
\lvert J\rvert &= \begin{vmatrix} \tfrac{\partial u}{\partial r} & \tfrac{\partial v}{\partial r} \\
\tfrac{\partial u}{\partial \theta} & \tfrac{\partial v}{\partial \theta} \end{vmatrix} 
= \begin{vmatrix} \cos \theta & \sin \theta \\ -r \sin\theta & r\cos\theta \end{vmatrix} = r.
\end{align}

We can rewrite Eq.~\eqref{eq:joint_distr_1} as
\begin{align}
f_{r, \theta}(r, \theta)
&= \tfrac{r}{\pi\sigma^2}e^{-\tfrac{\mu_r^2}{\sigma^2} \sin^2 (\mu_\theta - \theta)}e^{-\left(\tfrac{r - \mu_r \cos (\mu_\theta - \theta)}{\sigma^2}\right)^2}, \label{eq:joint_distr_2}
\end{align}
where $\mu_r = \sqrt{\mu_u^2 + \mu_v^2}$, and $\mu_{\theta} = \tan^{-1}(\mu_v/\mu_u)$.

The marginal distribution for $\theta$ is $f_\theta(\theta) = \int_0^{\infty} f_{r, \theta}(r, \theta) dr.$
This integral is solved by substitution with $t = \tfrac{r - \mu_{\ssub{R}} \cos (\mu_\Theta - \theta)}{\sigma}$, which leads to $r = t\sigma + \mu_{\ssub{R}} \cos (\mu_\theta - \theta)$, $dt = dr/\sigma$, and $dr = \sigma dt$. Hence, we have
\begin{align}
f_\theta(\theta) &= \tfrac{1}{\pi}  e^{\!-\gamma\sin^2 (\mu_\theta - \theta)}\!\!\!
\int_{t|_{r=0}}^{\infty} \!\!\!\!\! \left(t \!+\! \sqrt{\gamma} \cos (\mu_\theta \!-\! \theta) \right)e^{-t^2} dt, \label{eq:marginal_distr_3}
\end{align}
where $\gamma = \tfrac{\mu_{\ssub{R}}^2}{\sigma^2}$ is the \gls{snr}, and from which we obtain
\begin{align}
f_\theta(\theta) = &
\tfrac{1}{2\pi} e^{-\gamma}  +  \tfrac{\sqrt{\gamma}}{2\sqrt{\pi}}\cos (\mu_\theta \!-\! \theta) e^{-\gamma\sin^2 (\mu_\theta - \theta)}  \nonumber \\ & \times \Big[ 
 1 + \erf\left(\sqrt{\gamma}\cos (\mu_\theta \!-\! \theta)\right)\Big], \label{eq:marginal_distr_4}
\end{align}
with $\erf(\cdot)$ denoting the error function.

If $\gamma$ is large, i.e., $\mu_{\ssub{R}} \! \gg \! \sigma$, we can consider $e^{-\gamma} \! \approx \!0$, $\cos (\mu_\theta \!- \!\theta)e^{-\gamma \sin^2 (\mu_\theta - \theta)} \!\approx\! e^{-\gamma(\mu_\theta - \theta)^2}$, and $\erf\left(\sqrt{\gamma}\cos (\mu_\theta \!-\! \theta)\right) \!\approx\! 1$, allowing us to approximate Eq.~\eqref{eq:marginal_distr_4} as follows
\begin{align}
f_\theta(\theta) & \approx \tfrac{\sqrt{\gamma}}{\sqrt{\pi}} e^{-\gamma(\mu_\theta - \theta)^2}  = \tfrac{1}{\sqrt{2\pi}\sigma_{\theta}} e^{-\tfrac{(\mu_\theta - \theta)^2}{2\sigma_{\theta}^2}}, \label{eq:gaussian_approx}
\end{align}
with $\sigma_{\theta}=\tfrac{1}{\sqrt{2\gamma}}$, which proves the validity of Eq.~\eqref{eq:theta_distribution}.
\end{appendices}
\bibliographystyle{IEEEtran}
\bibliography{references}

\vskip -2.5\baselineskip plus -1fil

\begin{IEEEbiography}
[{\includegraphics[width=1in,height=1.25in,clip,keepaspectratio]{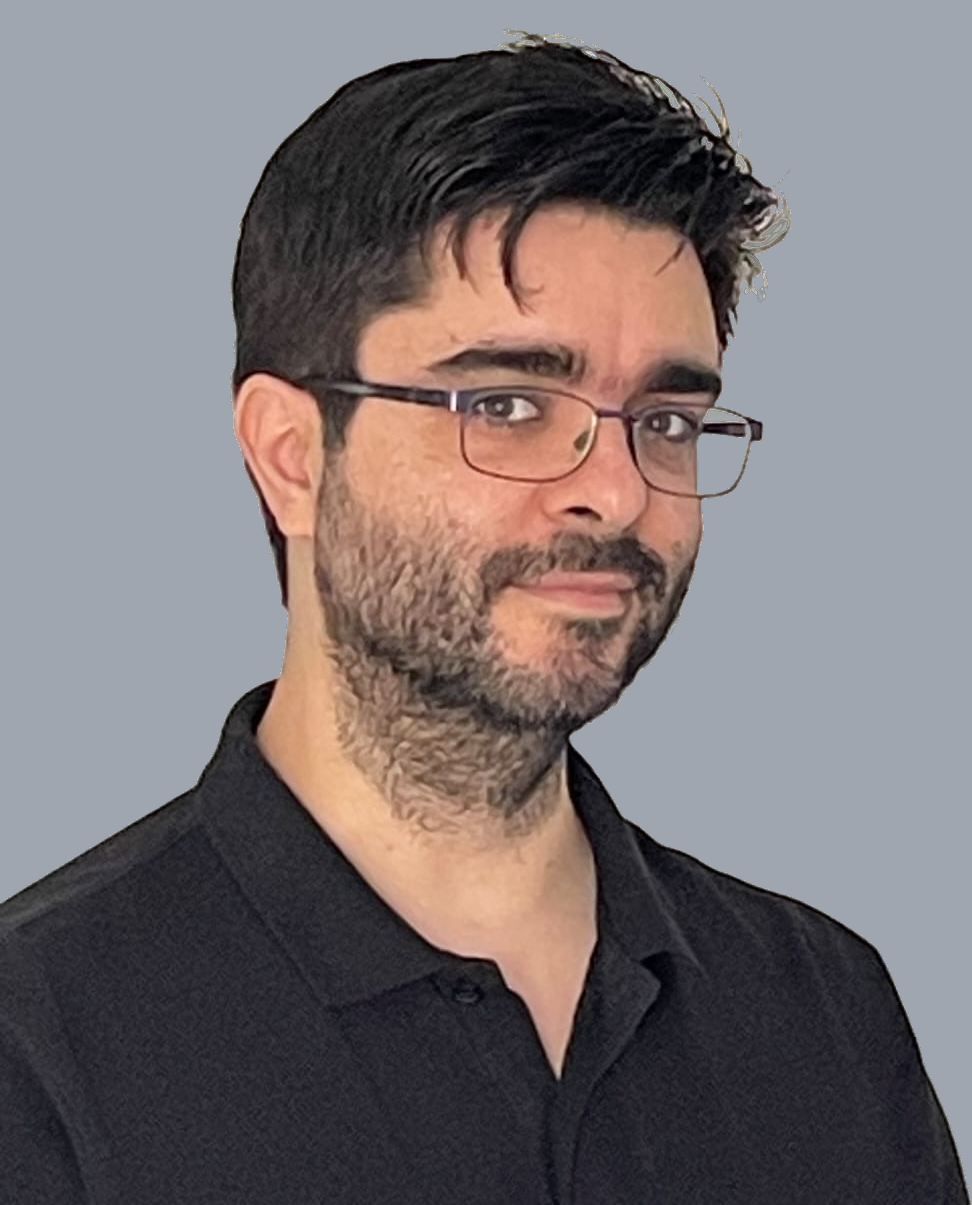}}]
{Guillermo Encinas-Lago } (M'21) received his M.Sc. degrees in Applied Physics from Universidad Autónoma de Madrid in 2013 and in Industry 4.0 from Univesidad Carlos III de Madrid in 2021, both in Spain, and a Ph.D. in Telecommunications in 2024 from Université Paris-Saclay, France, while employed at NEC Laboratories Europe in Heidelberg, Germany, under the MSCA program. He works now as Senior Researcher in i2Cat, focused on RISs, machine learning, ISAC, and prototyping.
\end{IEEEbiography}

\vskip -2.5\baselineskip plus -1fil

\begin{IEEEbiography}
[{\includegraphics[width=1in,height=1.25in,clip,keepaspectratio]{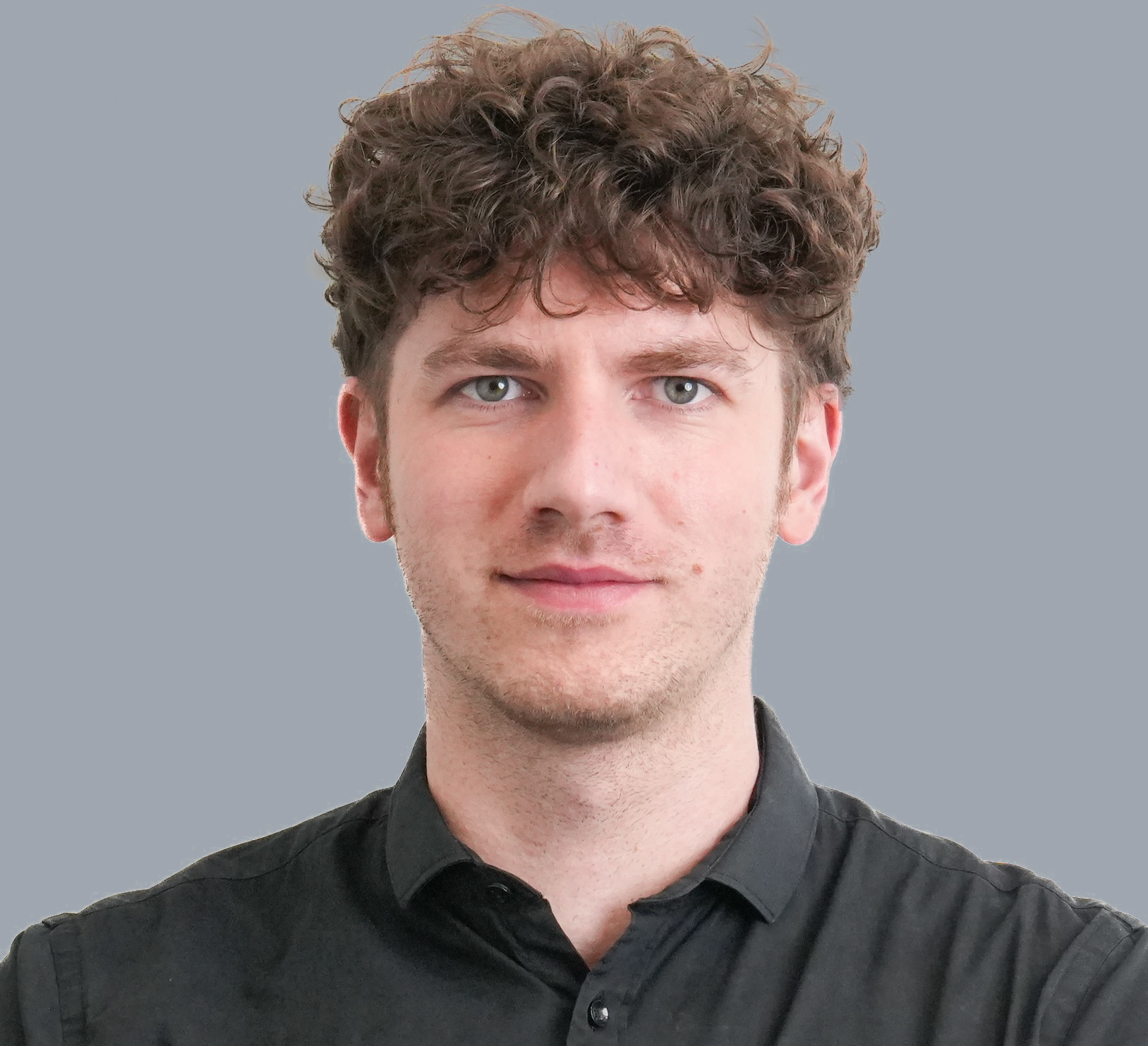}}]
{Francesco Devoti} (M’20) is a Senior Research Scientist at NEC Laboratories Europe GmbH. His interests are mobile radio networks, RISs, and ISAC, with a focus on mathematical modeling and optimization strategies.  He has been involved in several published international research papers and patents. He earned his B.S. and M.S. degrees in Telecommunication Engineering and a Ph.D. from the Politecnico di Milano.
\end{IEEEbiography}

\vskip -2.5\baselineskip plus -1fil

\begin{IEEEbiography}
[{\includegraphics[width=1in,height=1.25in,clip,keepaspectratio]{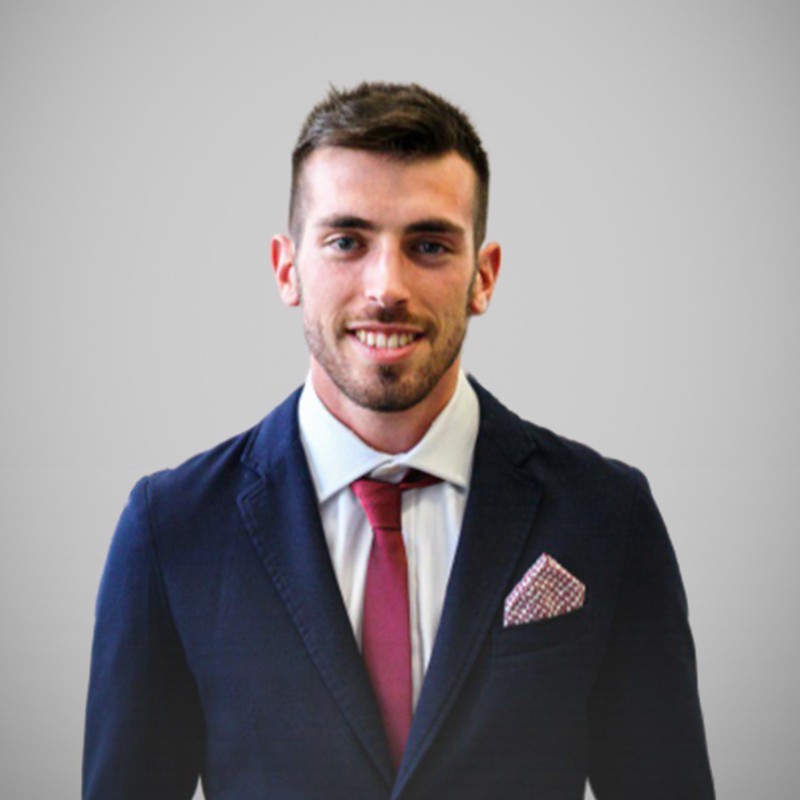}}]
{Marco Rossanese} (M'21) received the B.Sc. and M.Sc. degrees in Telecommunication Engineering from Università degli Studi di Padova in 2017 and 2019, respectively. He received a Ph.D in computer science from the Technische Universität of Darmstadt. He is employed as a researcher at NEC Laboratories Europe GmbH in the 6G Networks team. He focuses his work on RISs.
\end{IEEEbiography}

\vskip -2.5\baselineskip plus -1fil

\begin{IEEEbiography}
[{\includegraphics[width=1in,height=1.25in,clip,keepaspectratio]{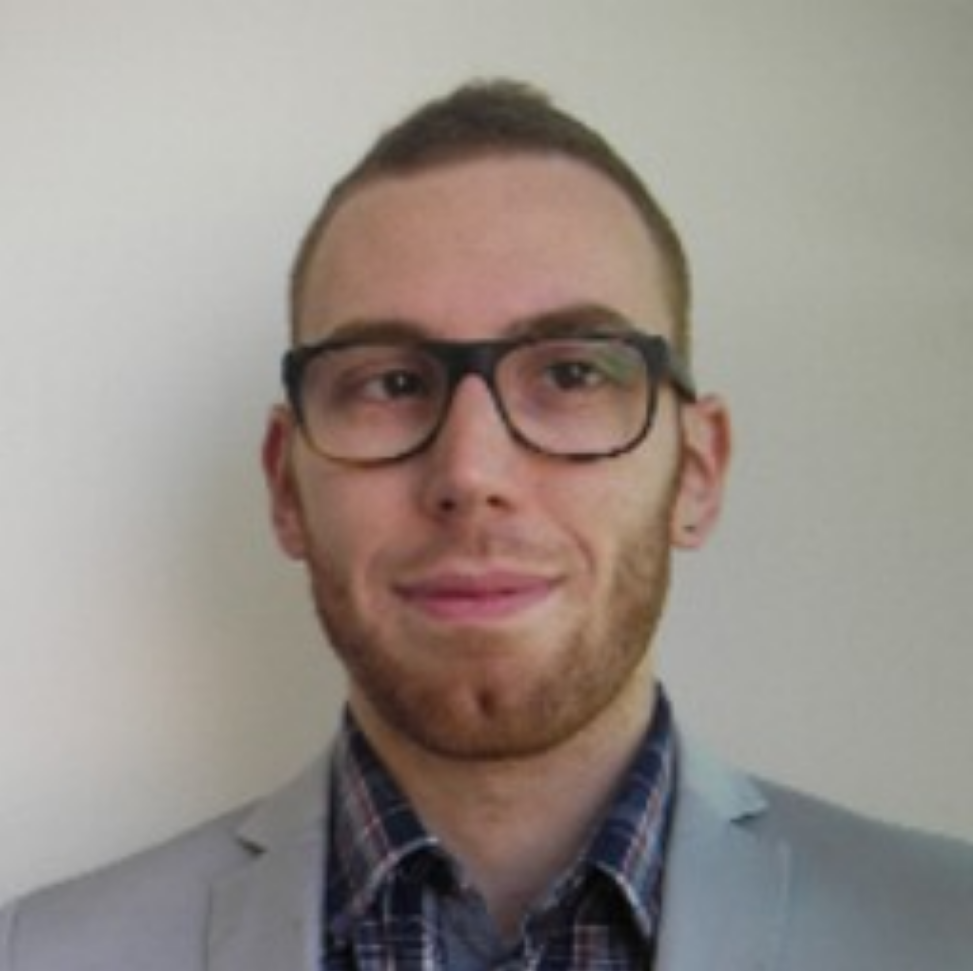}}]
{Vincenzo Sciancalepore} (M'15--SM'19) received his M.Sc. degree in Telecommunications Engineering and Telematics Engineering in 2011 and 2012. In 2015, he received a double Ph.D. degree. He is a Principal Researcher at NEC Laboratories Europe, focusing his activity on 6G RAN technologies, such as RISs. He is an editor of IEEE TCOM and IEEE TWC.
 \end{IEEEbiography}

\vskip -2.5\baselineskip plus -1fil

\begin{IEEEbiography}
[{\includegraphics[width=1in,height=1.25in,clip,keepaspectratio]{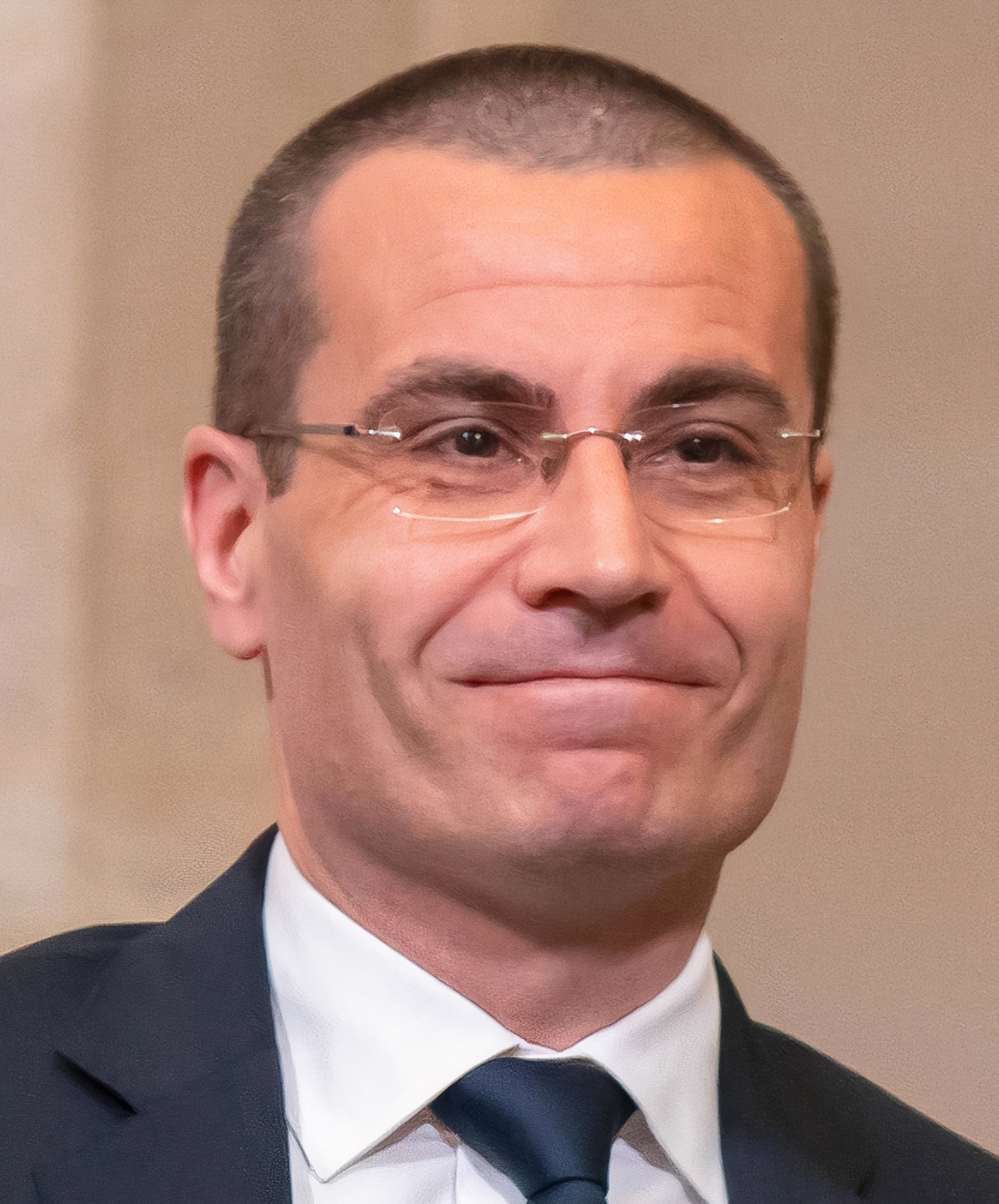}}]
{Marco Di Renzo} (Fellow, IEEE) received the Ph.D. degree in electrical engineering from the University of L’Aquila, Italy, in 2007. Currently, he is a CNRS Research Director (Professor) and the Head of the Intelligent Physical Communications group with the Laboratory of Signals and Systems (L2S) at CNRS \& CentraleSupélec, Paris-Saclay University, Paris, France, as well as a Chair Professor in Telecommunications Engineering with the Centre for Telecommunications Research -- Department of Engineering, King’s College London, London, United Kingdom. He served as the Editor-in-Chief of IEEE Communications Letters, and he currently serves as the Director of Journals of the IEEE Communications Society.
\end{IEEEbiography}

\vskip -2.5\baselineskip plus -1fil

\begin{IEEEbiography}
[{\includegraphics[width=1in,height=1.25in,clip,keepaspectratio]{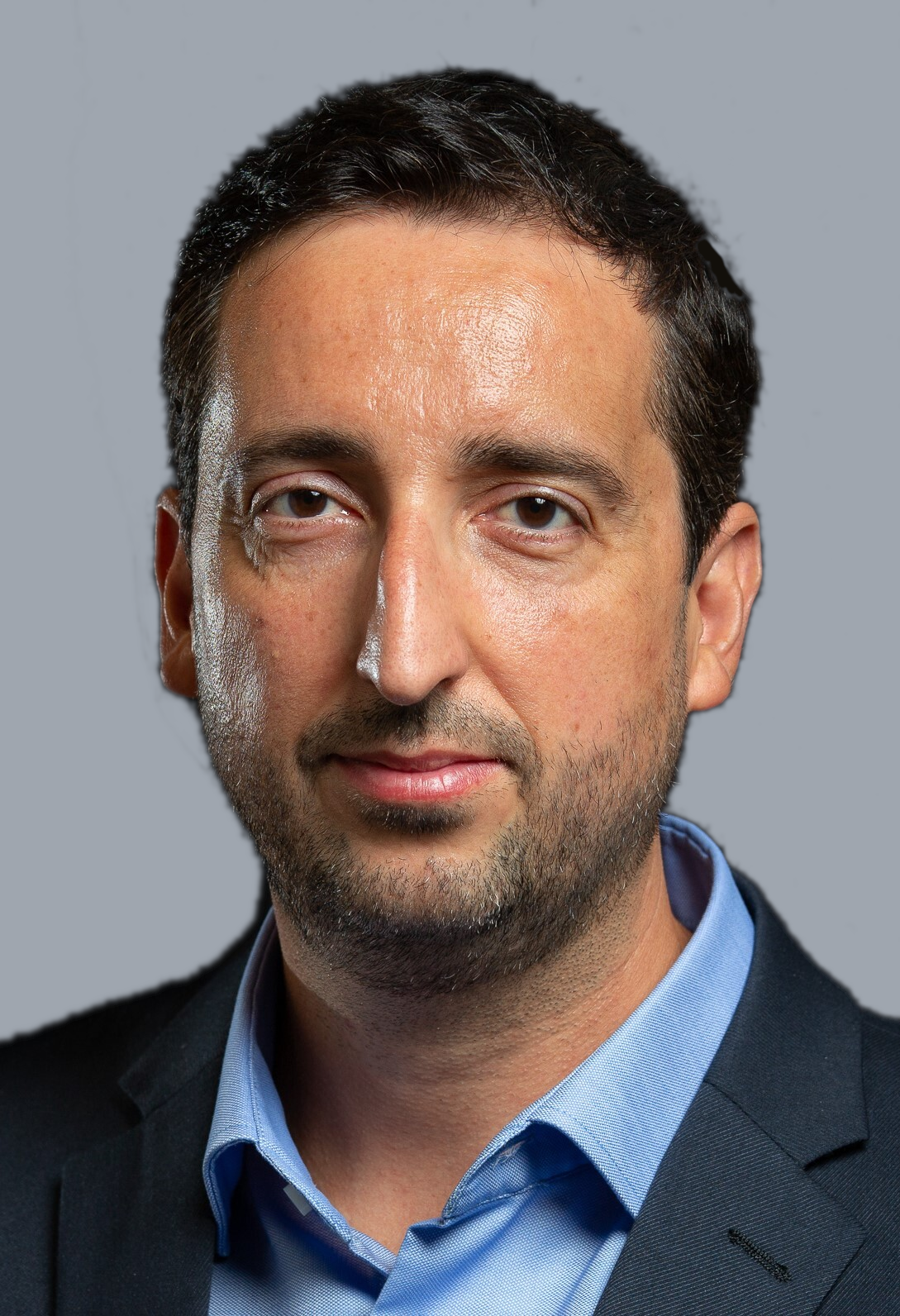}}]
{Xavier Costa-P\'erez} (M'06--SM'18) is ICREA Research Professor, Scientific Director at the i2Cat Research Center and Head of 5G/6G Networks R\&D at NEC Laboratories Europe. He has served on the Organizing Committees of several conferences, published papers of high impact and holds tenths of patents. Xavier received  his  Ph.D. degree in Telecommunications from the Polytechnic University of Catalonia.
\end{IEEEbiography}

\end{document}